\documentclass[12pt]{article}
\usepackage{amsfonts,amsthm,amsmath,amssymb,upgreek}
\usepackage[paper=letterpaper,margin=1in]{geometry}
\usepackage[export]{adjustbox}
\usepackage[numbers,sort&compress]{natbib}
\usepackage{color}
\usepackage{graphicx}
\usepackage{dcolumn}
\usepackage{bm}
\usepackage{hyperref}
\usepackage{float}
\usepackage{makecell}
\usepackage[caption=false]{subfig}
\usepackage{hyperref}
\hypersetup{
    colorlinks=true,
    linktoc=page,    
    linkcolor=blue,
    urlcolor=magenta
}
\usepackage{tikz,mathrsfs}
\usetikzlibrary{decorations.pathmorphing, decorations.markings, arrows.meta}
\usetikzlibrary{shadings}
\newcommand{\be}{\begin{equation}}
\newcommand{\ee}{\end{equation}}
\newcommand{\bea}{\begin{eqnarray}}
\newcommand{\eea}{\end{eqnarray}}
\newcommand{\dd}{\text{d}}

\newcommand{\tob}{\theta_{\text{ob}}}

\newcommand{\R}{\mathcal{R}}

\newcommand{\daggerfootnote}[1]{%
    \renewcommand{\thefootnote}{\fnsymbol{footnote}}%
    \footnote[2]{#1}
    \renewcommand{\thefootnote}{\arabic{footnote}}%
}

\begin{document}

\thispagestyle{empty}

\vspace*{.5cm}
\begin{center}

{\bf {\Large On the Cuspy Structure of Rotating Wormhole Shadows}\\
\vspace{1cm}}
    
 {\bf Peng Cheng\daggerfootnote{p.cheng.nl@outlook.com}, Ruo-Fan Xu, and Peng Zhao}\\
  \bigskip \rm
  
\bigskip

Center for Joint Quantum Studies and Department of Physics, \\School of Science, Tianjin University, Tianjin 300350, China\\

\rm

\vspace{2.5cm}
{\bf Abstract}
\end{center}
\begin{quotation}

We investigate the shadow cast by a rotating traversable wormhole in the Teo class endowed with a general redshift function, with particular emphasis on the emergence of cuspy structures. The shadow boundary is the common envelope of two critical orbit families: unstable circular orbits outside the throat and orbits at the throat itself. 
The formation of cusps, marking the transition between smooth and cuspy shadow boundaries, only becomes possible when the redshift parameter $\lambda$ is allowed to vary.
Moreover, we uncover a universal critical value $\lambda_c=(\sqrt5-1)/4$ that signals the onset of the cusp. 
A phase diagram characterized by the spin and redshift parameters reveals four distinct morphologies: smooth, cuspy, ears touching, and throat drowning.
The morphology of the wormhole shadow may provide observational diagnostics for the different compact objects in future high-resolution imaging observations.

\end{quotation}

\vspace{1cm}

\setcounter{page}{0}
\setcounter{tocdepth}{2}
\setcounter{footnote}{0}


\newpage
{\noindent} \rule[-10pt]{16.5cm}{0.05em}\\
\tableofcontents
{\noindent} \rule[-10pt]{16.5cm}{0.05em}\\


\section{Introduction}
\label{intro}

The shadow of a compact gravitating object has emerged as a cornerstone of strong-field gravity phenomenology. 
Since the Event Horizon Telescope's groundbreaking images of M87* and Sagittarius A*, the study of black hole shadows has matured into a precise and rich discipline, offering important insight into spacetime geometry of ultracompact objects \cite{EventHorizonTelescope:2019ggy,EventHorizonTelescope:2022wxl,EventHorizonTelescope:2021dqv,Gralla:2019xty,Cunha:2018acu,Perlick:2021aok}.  
Yet black holes are not the only theoretically viable compact objects. More general geometries, such as naked singularities, gravastars, and wormholes, can also cast observable shadows \cite{Sakai:2014ypa,Cunha:2018uzx,Nedkova:2013msa,Shaikh:2018kfv,Gyulchev:2018fmd}.
In particular, as one of the most intriguing predictions of Einstein's theory of gravity \cite{Einstein:1935rr,Wheeler:1955zz}, traversable wormholes, which constitute spacetime tunnels linking remote parts of the universe, are both rigorously formulated scientific models and enduring subjects of theoretical investigation.
The observational signatures, particularly their shadows, provide a concrete pathway to test the black hole paradigm and to search for horizonless alternatives \cite{DeSimone:2025sgu,DeFalco:2020afv,DeFalco:2021klh,DeFalco:2021btn,DeFalco:2023kqy,Dai:2019mse,Simonetti:2020ivl,Bambi:2021qfo,Amir:2018pcu,Kasuya:2021cpk,Tsukamoto:2020uay,Tsukamoto:2023rzd,Paul:2019trt}.
Thus, the investigation of the shadow cast by traversable wormholes holds significant importance.

Traversable wormholes were originally conceived by Morris and Thorne as solutions of Einstein's gravity coupled to exotic matter \cite{Morris:1988zz}.
To construct a singularity-free metric to describe wormhole geometry that is consistent with the Einstein field equations, the matter contained in Morris-Thorne wormholes must possess a stress-energy tensor that violates the null energy condition. 
Subsequently, rotating axially symmetric solutions describing traversable wormholes were also obtained \cite{Teo:1998dp,Kuhfittig:2003xt}. 
The apparent violation of the energy conditions may be either mitigated or circumvented depending on the theoretical context \cite{Visser:1989kg,Sushkov:1992tb,Hochberg:1996wp,Lobo:2005us,Sushkov:2005kj,Kanti:2011zz,Dai:2020rnc,Dai:2018vrw,Cisterna:2023uqf}, which establishes the traversable wormhole as a robust and conceptually rich model.
In this work, we will study the shadow of a general class of rotating traversable wormholes in the Teo class \cite{Teo:1998dp}.

A further reason for studying the shadow cast by rotating wormholes is due to the increasing attention on the cuspy shadow boundaries \cite{Wei:2026zwu,Wang:2017hjl,Chen:2025jay,Qian:2021qow,Cunha:2017qae,Held:2019xzy}.
The cusp is not numerical noise, but a distinctive structure dictated by the photon orbits in a strong gravitational system.
The recent study \cite{Wei:2026zwu} reveals that the formation of cusps in black hole shadows constitutes a topological phase transition and a gravitational equal-area law. Moreover, the critical point of cusp formation has universal exponents, placing the phenomenon in the mean-field universality class. 
The study implies that the emergence of cuspy structures is endowed with deep physical meaning, rendering them theoretically significant.
The shadow of the rotating traversable wormhole has also been reported to exhibit a cuspy edge \cite{Shaikh:2018kfv,Gyulchev:2018fmd}.
Thus, we are naturally eager to comprehensively understand the cuspy structure of the wormhole shadow.
However, early investigations of rotating wormhole shadows do not provide a careful enough study on the formation of the cuspy edge \cite{Nedkova:2013msa,Shaikh:2018kfv,Gyulchev:2018fmd,Abdujabbarov:2016efm,Harko:2009xf,Bambi:2013jda,Bronnikov:2021liv}.
Especially, most of these studies have adopted the simplest possible redshift function, typically $N=\exp⁡(-r_0/r)$ or $N=\exp⁡(-r_0^2/r^2)$. 
As will be illustrated in this work, those models are too simple to properly capture the process of cusp formation. 
In principle, the redshift function $N(r)$ can be chosen more freely, like 
\begin{equation}
	N = \exp\left[-\frac{r_0}{r} - \lambda\left(\frac{r_0}{r}\right)^2\right]\,.\label{eq:N0}
\end{equation}
We work with the above model and illustrate the underlying physics for the cusp formation.
We will investigate how, and to what extent, the redshift profile influences the shadow morphology.
Note that the parameter $\lambda$ controls the steepness of the redshift function as one approaches the throat from large $r$.
For very small $\lambda$, we recover the familiar $N = \exp(-r_0/r)$ model, whereas the redshift function becomes steeper as one increases $\lambda$. 
For very large $\lambda$, the function is dominated by the term $\exp(-\lambda \cdot r_0^2/r^2)$, leading to a much faster decay as $r$ decreases.

As will be demonstrated in this paper, the shadow of rotating traversable wormholes exhibits a surprisingly rich structure governed by the redshift parameter $\lambda$ and spin parameter $a$. More explicitly, we uncover a universal critical value $\lambda_c=(\sqrt{5}-1)/4$ independent of spin and throat radius, below which the shadow boundary is smooth and above which a sharp cusp emerges at the intersection of the outer circular orbits and the throat critical orbits. 
The formation of cusps can not be seen by the models with $N=\exp⁡(-r_0/r)$ or $N=\exp⁡(-r_0^2/r^2)$ \cite{Nedkova:2013msa,Shaikh:2018kfv,Gyulchev:2018fmd}.
By scanning the $(\lambda, a)$ parameter space, a comprehensive phase diagram is constructed, revealing four distinct shadow morphologies: smooth, cuspy, ears touching, and throat drowning.
As we vary the parameters, the shadow can exhibit interesting re-entrant behavior where the curve of the throat orbits detaches and later reconnects. The cusp boundary enables a better understanding of the photon orbits in the wormhole background, and the special morphology may serve as a diagnostic signature for distinguishing traversable wormholes from black holes and other compact objects.

The paper is organized as follows.
Sec. \ref{sec:metric} introduces the rotating traversable wormhole metric and specifies the class of solutions studied. 
We derive the null geodesic equations, the conditions for critical orbits, and the mapping to the observer's celestial coordinates in Sec. \ref{sec:shadow}. Then in Sec. \ref{sec:cusp}, we investigate the shadow morphology with different spin and redshift parameters, revealing a universal critical redshift parameter​ and a phase diagram comprising four distinct phases. Sec. \ref{sec:con} concludes the whole paper and discusses future directions. 
Appendix \ref{sec:para} provides a parametric description of the wormhole throat orbits.


\section{Rotating traversable wormholes}
\label{sec:metric}

The metric of a rotating traversable wormhole is given in the Teo form as \cite{Teo:1998dp}
\begin{equation}
ds^{2} = -N^{2}\dd t^{2} + \left(1 - \frac{b}{r}\right)^{-1}\dd r^{2} + r^{2}K^{2}\left[\dd \theta^{2} + \sin^{2}\theta (\dd \phi - \omega \dd t)^{2}\right],\label{metric}
\end{equation}
where $N, b, K, \omega$ are functions of the radial and polar coordinates $r$ and $\theta$. 
We require these functions to be independent of $\theta$ to avoid possible singularities at $\theta=0$ or $\pi$. Moreover, the geodesic equations on the geometry would be easier to deal with.
We now discuss the physical meaning of these metric functions in more detail
\begin{itemize}
  \item The redshift function $N$ determines the gravitational redshift. To ensure the wormhole is traversable, the redshift function $N$ should be finite and nonzero. $N(r)$ is usually chosen to be the form frequently used in the literature \cite{Nedkova:2013msa,Shaikh:2018kfv,Gyulchev:2018fmd,Abdujabbarov:2016efm,Harko:2009xf,Bambi:2013jda}
\begin{equation}
	N = \exp\left[-\frac{r_0}{r} - \lambda\left(\frac{r_0}{r}\right)^\delta\right]\label{eq:N}
\end{equation}
with parameters $r_0$, $\lambda$, and $\delta$.
For concrete calculations, we will fix $\delta=2$, while keeping $\lambda$ as a free parameter. 
\item The shape function $b$ determines the shape of the wormhole. As can be seen from \eqref{metric}, the metric contains an apparent singularity at $r=b\geq 0$, which corresponds to the throat of the wormhole. According to Morris and Thorne's flare out condition \cite{Morris:1988zz}, to have the characteristic ``flaring'' shape, the shape function must satisfy $ \partial_r b(r) < 1$ at the throat. The shape function is usually chosen to be
  \begin{equation}
	b(r)=r_0\left(\frac{r_0}{r}\right)^{\gamma}\,.
\end{equation}
When $\gamma=0$, the throat is located at $r=r_0$, and we have $r_0=GM$, with $M$ being the ADM mass of the wormhole. This is the shape function we will deal with in this work.
\item For other metric functions, we adopt simple forms. The radial distance function $K$ determining the area radius will be set to $K=1$. $\omega$, measuring the angular velocity of the wormhole, can be related to the angular momentum $J$ by
\begin{equation}
	\omega=\frac{2J}{r^3}\,,\label{eq:omega}
\end{equation} 
and the spin parameter is defined as $a=J/M^2$.
\end{itemize}
The radial coordinate $r$ ranges from $r_0 \leq r < \infty$, where $r \to \infty$ corresponds to physical infinity. For physical relevance, the solution is assumed to be asymptotically flat.
Actually, apart from the restrictions necessary for regularity and physical reasonableness, the metric functions can be chosen freely. Each specific choice corresponds to a particular rotating traversable wormhole.
To concretely illustrate the main point of this work, we will work with the specific rotating traversable wormhole discussed in (\ref{eq:N}-\ref{eq:omega}).
So, as a brief summary, the remainder of this work focuses on a special class of solutions where the metric functions adopt simple forms 
\begin{equation}
N = \exp\left[-\frac{r_0}{r} - \lambda\left(\frac{r_0}{r}\right)^2\right]\,,\quad b = r_0,\quad K = 1,\quad \omega = \frac{2J}{r^{3}}.\label{metricF}
\end{equation}
In the limit of zero angular velocity ($\omega = 0$), this class of solutions reduces to the classic static Morris-Thorne wormhole \cite{Morris:1988zz}.

\section{Shadow of the wormholes}
\label{sec:shadow}

With the rotating traversable wormhole geometry \eqref{metric}, we can follow the standard protocol to work out the shadow of those rotating traversable wormholes.
We will provide a detailed review of the wormhole shadow measured by a distant observer in this section. 

\subsubsection*{Null geodesics}

The Lagrangian describing null geodesics in the rotating traversable wormhole geometry \eqref{metric} can be expressed as
\begin{equation}
	\mathcal{L}=\frac{1}{2}g_{\mu\nu}\frac{\dd x^\mu}{\dd \sigma}\frac{\dd x^\nu}{\dd \sigma}\,,\label{lag}
\end{equation}
where the Lagrangian is defined with respect to the affine parameter $\sigma$.
The corresponding momentum for the particle can be derived using the Lagrangian
\begin{equation}
	p_\mu=\frac{\partial \mathcal{L}}{\partial \dot{x}^{\mu}}\,,\label{pmu}
\end{equation}
where $\dot{x}^{\mu}$ is $\dd x^{\mu}/\dd \sigma$. Because the Lagrangian $\mathcal{L}$ is independent of $t$ and $\phi$, there are two conserved quantities: conserved energy $E$ and angular momentum $L$
\begin{equation}
	\begin{split}
		E &=-p_t= N^2 \frac{\dd t}{\dd \sigma} + \omega r^2 K^2 \sin^2\theta \left( \frac{\dd \phi}{\dd \sigma} - \omega \frac{\dd t}{\dd \sigma} \right),\\
		L &=p_\phi= r^2 K^2 \sin^2\theta \left( \frac{\dd \phi}{\dd \sigma} - \omega \frac{\dd t}{\dd \sigma} \right).
	\end{split}\label{eq:EL}
\end{equation}
The Hamiltonian $\mathcal{H}$ associated with \eqref{lag} is
\begin{equation}
	\mathcal{H}=\frac{1}{2}p_{\mu} p^{\mu}\,,
\end{equation}
The Hamilton-Jacobi equation
\begin{equation}
	\frac{\partial \mathcal{S}}{\partial \sigma}+\mathcal{H} =0
\end{equation}
can be rewritten as the following equation
\begin{equation}
\frac{\partial \mathcal{S}}{\partial \sigma} = -\frac{1}{2} g^{\mu\nu} \frac{\partial \mathcal{S}}{\partial x^\mu} \frac{\partial \mathcal{S}}{\partial x^\nu}\label{HJ2}
\end{equation}
where $\mathcal{S}$ is the Jacobi action
\begin{equation}
\mathcal{S} = \frac{1}{2} \mu^2 \sigma - E t + L \phi + \mathcal{S}_r(r) + \mathcal{S}_\theta(\theta)
\end{equation}
For a photon, the mass $\mu = 0$. 
Substituting the metric, we can rewrite equation \eqref{HJ2} as
\begin{equation}
\left(1 - \frac{b}{r}\right) N^2 \left( \frac{\partial\mathcal{S}_r}{\partial r} \right)^2 - (E - \omega L)^2
= -\frac{N^2}{r^2 K^2} \Bigg[ \left( \frac{\partial\mathcal{S}_\theta}{\partial\theta} \right)^2 + \frac{L^2}{\sin^2\theta} \Bigg].\label{eq:sep}
\end{equation}
The left-hand side depends only on $r$, while the right-hand side depends on $\theta$ multiplied by the common factor $-\frac{N^2}{r^2 K^2}$. For the equality to hold for all $r$ and $\theta$, both sides must equal a constant. We use the Carter constant $-\mathcal{K}$ to denote the part inside the square bracket in \eqref{eq:sep}.
 Therefore, we have
\begin{equation}
\begin{split}
	\left( \frac{\partial\mathcal{S}_\theta}{\partial \theta} \right)^2 &= \mathcal{K} - \frac{L^2}{\sin^2\theta} \\
	 \left( 1 - \frac{b}{r} \right) N^2 \left( \frac{\partial\mathcal{S}_r}{\partial r} \right)^2 &= (E - \omega L)^2 -\mathcal{K} \frac{N^2}{r^2 K^2} \,,
\end{split}\label{sepHJ}
\end{equation}
According to the Hamilton-Jacobi method and the definition of $p_\mu$, we have
\begin{equation}
	\frac{\partial \mathcal{S}}{\partial x^\mu}=p_\mu=g_{\mu\nu} \frac{\dd x^\nu}{\dd \sigma}\,.
\end{equation}
Thus, we can use $\dd x^\mu/\dd \sigma$ to replace $\partial \mathcal{S}/\partial x^\mu$ in equation \eqref{sepHJ}.
Introducing the impact parameters
\begin{equation}
\xi = \frac{L}{E}, \qquad \eta = \frac{\mathcal{K}}{E^2} \,,
\end{equation}
we obtain the radial and polar geodesic equations for light as
\begin{equation}
\frac{N}{E}\left(1 - \frac{b}{r}\right)^{-1/2} \frac{\dd r}{\dd\sigma} = \pm \sqrt{\R(r)}, \qquad
\frac{r^2 K^2}{E} \frac{\dd\theta}{\dd\sigma} = \pm \sqrt{\Theta(\theta)}\,,\label{eq:geos}
\end{equation}
with 
\begin{equation}
\R(r) = (1 - \omega \xi)^2 - \eta \frac{N^2}{r^2 K^2}, \qquad
\Theta(\theta) =\eta - \frac{\xi^2}{\sin^2\theta}\,. \label{eq:RT}
\end{equation}
There are two more equations that can be derived from the conserved quantities demonstrated in equation \eqref{eq:EL}.
\begin{equation}
\frac{\dd t}{\dd \sigma} = \frac{E - \omega L}{N^2}, \qquad
\frac{\dd \phi}{\dd \sigma} = \frac{L}{r^2 K^2 \sin^2\theta} + \frac{\omega (E - \omega L)}{N^2} .
\end{equation}

Our primary focus is the radial geodesic equation, which can be recast as one-dimensional motion under an effective potential.
We rewrite the first equation in \eqref{eq:geos} as
\begin{equation}
\left( \frac{\dd r}{\dd \sigma} \right)^2 + V_{\text{eff}} = 0, \qquad
V_{\text{eff}} = -\frac{E^2}{N^2} \left( 1 - \frac{b}{r} \right) \R(r)\,,\label{Veff}
\end{equation}
where $V_{\text{eff}}$ is the effective potential describing the radial motion of the photon.

\subsubsection*{Two families of orbits}
  
The edge of the shadow corresponds to the critical orbits that separate escaping photons from those falling into the wormhole. 
The shadow consists of two parts: unstable circular orbits outside the throat and critical orbits at the throat \cite{Shaikh:2018kfv}.  The shadow is the common region enclosed by these two curves.
The critical orbits are unstable circular photon orbits satisfying the extremum conditions of the effective potential
\begin{equation}
V_{\text{eff}} = 0, \qquad
\frac{\dd V_{\text{eff}}}{\dd r} = 0, \qquad
\frac{\dd ^2 V_{\text{eff}}}{\dd r^2} \le 0 \,.\label{eq:edge}
\end{equation}

For unstable circular orbits away from the throat, conditions \eqref{eq:edge} are equivalent to
\begin{equation}
\R(r) = 0, \qquad
\frac{\dd \R}{\dd r} = 0, \qquad
\frac{\dd ^2\R}{\dd r^2} \ge 0 
\end{equation}
From $\R = 0$ and $\dd \R/\dd r = 0$, we solve for the impact parameters as functions of the circular orbit radius $r_{\text{ph}}$
\begin{equation}
\eta = \left[ \frac{r^2 K^2}{N^2} (1 - \omega \xi)^2 \right]_{r = r_{\text{ph}}} ,\qquad
\xi = \frac{\Sigma}{\Sigma \omega - \omega'} \bigg|_{r = r_{\text{ph}}}, \label{eta-xi}
\end{equation}
with
\begin{equation}
\Sigma = \frac{1}{2} \frac{\dd}{\dd r} \ln \left( \frac{N^2}{r^2 K^2} \right) \,.
\end{equation}
The prime in \eqref{eta-xi} denotes differentiation with respect to $r$.
Note that the $r_{\text{ph}}$ ranges from $r_0$ to infinity.

At the throat, because $1 - b/r = 0$ is satisfied, the extremum conditions for unstable orbits \eqref{eq:edge} should be expressed differently.
At $r=r_0$ the factor $(1-b/r)$ vanishes, so we always have $V_{\text{eff}}(r_0)=0$ irrespective of $\mathcal{R}(r_0)$. 
A straightforward calculation gives
\begin{equation}
\left.\frac{\dd V_{\text{eff}}}{\dd r}\right|_{r=r_0} = -\frac{\mathcal{R}(r_0)}{r_0 N(r_0)^2},
\end{equation}
\begin{equation}
\left.\frac{\dd ^2V_{\text{eff}}}{\dd r^2}\right|_{r=r_0} = -\frac{2\,\mathcal{R}'(r_0)}{r_0 N(r_0)^2},
\end{equation}
where we have used $\mathcal{R}(r_0)=0$ when evaluating the second derivative.  Hence the extremum conditions reduce to
\begin{equation}
\R(r_0) = 0, \qquad
\frac{\dd \R}{\dd r} \bigg|_{r = r_0} \ge 0\,,
\end{equation}
The first condition $\R(r_0) = 0$ gives
\begin{equation}
(1 - \omega_0 \xi)^2 - \eta \frac{N_0^2}{r_0^2 K_0^2} = 0 \label{throat}
\end{equation}
Here, the subscript ``0'' denotes evaluation at the throat $r = r_0$. The second determines its instability, i.e., $\frac{\dd \R}{\dd r}|_{r=r_0}>0$ corresponds to unstable orbits.

\subsubsection*{Shadows in the sky}

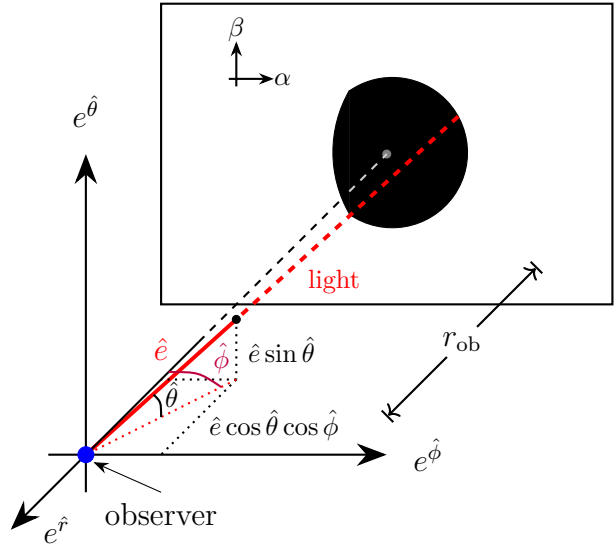
\begin{figure}[H]
\centering
\begin{tikzpicture}
 \node at (0,4) {
 	\begin{tikzpicture}[scale=1]
    \draw[thick] (0,0) rectangle (6,4);
	\draw [-{Stealth[length=2mm]}, thick] (1,3-0.1)--(1,3.5)  node[midway,above=0.15cm]{\footnotesize $\beta$};
	\draw [-{Stealth[length=2mm]}, thick] (1-0.1,3)--(1.5,3)  node[midway,right=0.15cm]{\footnotesize $\alpha$};
	\filldraw [fill=black] (2.505,2.85) arc (150:210:1.65);
	\filldraw [fill=black] (2.5,1.2) arc (-125:125:1);
	 \filldraw [black!50] (3,2) circle (1.5pt);
	\end{tikzpicture}};
 \draw [red,line width =1.5pt](-2,1.8) --(-4,0);
 \draw [dashed,red,line width =1.5pt](0.95,4.5) --(-2,1.8);
 \filldraw [black] (-2,1.8) circle (1.5pt);
 \draw [dotted,thick] (-2,1.8)--(-2,1);
 \draw [dotted,thick] (-2,1)--(-3,0);
 \draw [dotted,thick] (-2,1)--(-3,1);
 \draw [red,dotted,thick] (-2,1)--(-4,0);
 \draw [-{Stealth[length=4mm]}, thick] (-4.5,0)--(0,0)  node[right=0.2cm]{$e^{\hat{\phi}}$};
 \node [red] at (-0.7,2.3) {\footnotesize light};
 \node at (1,1.5) {$r_{\text{ob}}$};
 \draw [->|, thick] (1.2,1.7)--(2,2.5);
 \draw [->|, thick] (0.8,1.3)--(0,0.5);
 \draw [-{Stealth[length=4mm]}, thick] (-4,-0.5)--(-4,4)  node[above=0.2cm]{$e^{\hat{\theta}}$};
 \draw [-{Stealth[length=4mm]},thick] (-2.5,1.5)--(-5,-1) node[right=8pt]{$e^{\hat{r}}$};
 \draw [black!20,dashed,thick] (-0.5,3.5)--(0,4);
 \draw [black,dashed,thick] (-2.5,1.5)--(-0.5,3.5);
 \filldraw [blue] (-4,0) circle (3pt);
 \draw [{Stealth[length=2mm]}-](-3.9,-0.1) --(-3,-0.5)node[below] {observer};
 \draw [black,thick] (-3,0.5)..controls (-3,0.7)..(-3.1,0.8) node[right] {\footnotesize $\hat{\theta}$};
 \draw [purple,thick] (-2.9,1.1)..controls (-2.5,1.1)..(-2.2,0.9) node[above=1pt] {\footnotesize $\hat{\phi}$};
 \node at (-1.4,1.4) {\footnotesize $\hat{e}\sin \hat{\theta}$};
 \node at (-1.5,0.4) {\footnotesize $\hat{e}\cos \hat{\theta}\cos\hat{\phi}$};
 \node [red] at (-3,1.4) {$\hat{e}$};
\end{tikzpicture}
\caption{The shadow of a compact object in the observer's sky. Bardeen's impact parameters can be related to the observer's local frame $(\hat{r},\hat{\theta},\hat{\phi})$.}\label{fig:shadow}
\end{figure}

Let us consider the image of the rotating wormhole observed by a distant observer.
Assume the observer is located in one asymptotic region of the wormhole, far from the throat ($r_{\text{ob}} \to \infty$), and with an inclination angle $\theta_{\text{ob}}$ which is the angle between the observer's line of sight and the rotation axis of the wormhole. The equatorial plane is the plane with $\theta_{\text{ob}}=\pi/2$.
For the current problem under consideration, the metric approximates the Minkowski form in the asymptotic region.
Then, the local orthonormal frame for the observer can be regarded as a coordinate transformation
\begin{equation}
e_{\hat{t}} = \partial_t, \qquad
e_{\hat{r}} = \partial_r, \qquad
e_{\hat{\theta}} = \frac{1}{r} \partial_\theta, \qquad
e_{\hat{\phi}} = \frac{1}{r\sin\theta} \partial_\phi.\label{eq:local}
\end{equation}
We see that the asymptotic observer and the wormhole system are using the $(t,r)$ coordinates.
As illustrated in Fig. \ref{fig:shadow}, we can use $\hat{\theta}$ and $\hat{\phi}$ to denote the spherical coordinates of the asymptotic observer, which are related to the wormhole coordinates by a transformation shown in \eqref{eq:local}.

Consider a photon coming from the wormhole direction, as illustrated by the red line in Fig. \ref{fig:shadow}.
The photon's four-momentum can be written as
\begin{equation}
	p=p^{\mu}\partial_{\mu}=p^{\hat{\mu}}\partial_{\hat{\mu}}\,.
\end{equation}
In the local frame, we have
\begin{equation}
\begin{aligned}
p^{\hat{r}} &= |\vec{p}| \cdot \cos{\hat{\theta}}\cos{\hat{\phi}}, \\
p^{\hat{\theta}} &= |\vec{p}| \cdot \cos{\hat{\theta}}\sin{\hat{\phi}}, \\
p^{\hat{\phi}} &= |\vec{p}| \cdot \sin{\hat{\theta}}.
\end{aligned}\label{eq:relation}
\end{equation}
On the other hand, $p^{\hat{\mu}}$ can be related to the four-momentum $p^\mu$ defined in the wormhole coordinate.
The photon energy measured by the observer is $E_{\text{ob}} = -p \cdot e_{\hat{t}} = -p_t = E$.
The spatial components in the local frame are
\begin{equation}
\begin{aligned}
p^{\hat{r}} &= p \cdot e_{\hat{r}} = p^r, \\
p^{\hat{\theta}} &= p \cdot e_{\hat{\theta}} = r p^\theta, \\
p^{\hat{\phi}} &= p \cdot e_{\hat{\phi}} = r \sin\theta \, p^\phi.
\end{aligned}
\end{equation}

We can use Bardeen's impact parameters $(\alpha, \beta)$ to describe the apparent position of the shadow boundary in the observer's sky. 
As can be seen from Fig. \ref{fig:shadow}, Bardeen's impact parameters are related to the observer's local frame by
\begin{equation}
\alpha =-r\cdot \tan \hat{\phi},\qquad 
\beta = r\cdot \frac{\tan \hat{\theta}}{\cos \hat{\phi}}.\label{bardeen1}
\end{equation}
Note that the minus sign in \eqref{bardeen1} is due to the convention used in Bardeen's original definition \cite{Bardeen:1973tla}.
Substituting \eqref{eq:relation}, we can use $p^{\hat{\mu}}$ to express the local angles, and obtain
\begin{equation}
\begin{split}
	\alpha &= -r \cdot \frac{p^{\hat{\phi}}}{p^{\hat{r}}} = -r^2\sin\theta \cdot \frac{p^{\phi}}{p^{r}} ,\\
\beta &=  r  \cdot\frac{p^{\hat{\theta}}}{p^{\hat{r}}}= r^2  \cdot\frac{p^{\theta}}{p^{r}}.
\end{split}
\end{equation}
$p^{\mu}$ in the wormhole coordinates are expressed in \eqref{pmu}.
Specifically, we have
\begin{equation}
  \begin{split}
  	&p_\phi = \frac{L}{r^2 K^2 \sin^2\theta} + \frac{\omega (E - \omega L)}{N^2} \\
  	& p_r=  \frac{E}{N}\left(1 - \frac{b}{r}\right)^{1/2}\sqrt{\R(r)}, \\
  	& p_\theta= \pm \frac{E}{r^2 K^2}\sqrt{\Theta(\theta)}\,,
  \end{split}
\end{equation}
where $\R(r)$ and $\Theta(\theta)$ are expressed in \eqref{eq:RT}. Note that we take the photon moving outward from the wormhole to have $p_r>0$ (increasing $r$).
Therefore, for an observer located at infinity ($r=r_{\text{ob}} \to \infty$) with inclination angle $\theta_{\text{ob}}$, Bardeen's impact parameters $(\alpha, \beta)$ can be expressed as
 \begin{equation}
 \begin{split}
\alpha &= \lim_{\theta\to \theta_{\text{ob}}} \lim_{r\to\infty}\left( -r^2\sin\theta \cdot \frac{p^{\phi}}{p^{r}} \right)= -\frac{\xi}{\sin\theta_{\text{ob}}},\\[8pt]
\beta &=  \lim_{\theta\to \theta_{\text{ob}}} \lim_{r\to\infty}\left(r^{2} \cdot \frac{p^{\theta}}{p^{r}}\right)= \pm \sqrt{\eta - \frac{\xi^{2}}{\sin^{2}\theta_{\text{ob}}}}.
\end{split}
\end{equation}
This relation maps the impact parameter space $(\xi, \eta)$ to the observer's sky plane $(\alpha, \beta)$.

\subsubsection*{The shadow of rotating wormholes}

Now, we are ready to determine the wormhole shadow observed by the distant observer using $(\alpha, \beta)$.
As discussed earlier in this section, the boundary of the shadow consists of unstable circular orbits outside the throat and critical orbits at the throat. Let us look at the two sets of orbits separately.

\begin{itemize}
  \item \noindent\textbf{Critical circular orbits outside the throat}. For unstable circular orbits with radius $r_{\text{ph}} > r_0$, the conditions $\R(r)=0$ and $\dd\R/\dd r=0$ yield \eqref{eta-xi}.
Substituting $\xi$ and $\eta$ into the celestial coordinate expressions gives 
\begin{equation}
\begin{split}
	\alpha(r_{\text{ph}}) &= -\frac{1}{\sin\theta_{\text{ob}}} \left( \frac{\Sigma}{\Sigma \omega - \omega'} \right)\bigg|_{r=r_{\text{ph}}},\\[10pt]
\beta(r_{\text{ph}}) &= \pm \sqrt{ \left[ \frac{r^{2} K^{2}}{N^{2}} (1 - \omega \xi)^{2} - \frac{\xi^{2}}{\sin^{2}\theta_{\text{ob}}} \right] }\bigg|_{r=r_{\text{ph}}}.
\end{split}
\end{equation}
Working with the metric functions \eqref{metricF}, we obtain a parametric curve
\begin{equation}
	\begin{split}
		\alpha(r_{\text{ph}}) &= -\frac{1}{\sin\theta_{\text{ob}}} \cdot \frac{r^3 (r_0 r - r^2 + 2\lambda r_0^2)}{2J (r_0 r + 2 r^2 + 2\lambda r_0^2)}\bigg|_{r=r_{\text{ph}}}, \\[6pt]
\beta(r_{\text{ph}}) &= \pm \frac{r^3\sqrt{ 36 J^2 \sin^2\theta_{\text{ob}} \, N^{-2}(r) - (r_0 r - r^2 + 2\lambda r_0^2)^2 }}{2J (r_0 r + 2 r^2 + 2\lambda r_0^2) \sin\theta_{\text{ob}}} \bigg|_{r=r_{\text{ph}}}.
	\end{split}\label{eq:ab}
\end{equation}
$r_{\text{ph}}\in [r_0,\infty)$ is the running parameter characterizing unstable circular orbits radius.

\item \noindent\textbf{Critical orbits at the throat}.
At the throat $r = r_0$, the effective potential can exhibit an extremum. The critical conditions yield \eqref{throat}.
Combining with the celestial coordinate expressions to eliminate $\xi$ and $\eta$ yields a quadratic curve in $\alpha$ and $\beta$:
\begin{equation}
(N_0^{2} - \omega_0^{2} r_0^{2} K_0^{2} \sin^{2} \theta_{\text{ob}}) \alpha^{2} - 2\omega_0 r_0^{2} K_0^{2} \sin \theta_{\text{ob}} \, \alpha - r_0^{2} K_0^{2} + N_0^{2} \beta^{2} = 0.\label{eq:throat37}
\end{equation}
At the throat, we have
\begin{equation}
	N_0 = \exp\left[-\frac{r_0}{r_0} - \lambda\left(\frac{r_0}{r_0}\right)^2\right] = e^{-1-\lambda}, \quad K_0 = 1, \quad \omega_0 = \frac{2J}{r_0^3}.
\end{equation}
Thus, we can solve \eqref{eq:throat37} and get
\begin{equation}
	\beta = \pm N_0^{-1}\sqrt{ r_0^2\left(1+\omega_0 \sin\theta_{\text{ob}}\alpha\right)^2-\alpha^2 N_0^2}.\label{eq:throat39}
\end{equation}
The critical orbits at the throat are visible if the parameter $\alpha$ belongs to
\begin{equation}
	-\frac{r_0}{N_0+r_0\omega_0\cdot \sin \theta_{\text{ob}} }\leq \alpha\leq -\frac{2\lambda}{\left(2 \lambda+3 \right)\omega_0 \cdot\sin \theta_{\text{ob}}}\,.
\end{equation}
We have an explicit expression for the Bardeen impact parameter $(\alpha,\beta)$ for the throat orbits as displayed in \eqref{eq:throat39}, however, it may be useful to have a parametric expression for the throat orbits just like \eqref{eq:ab} as well. We discuss the parametric expression for the throat orbits in Appendix \ref{sec:para} by introducing an effective radius $\bar{r} \in [0,r0]$.
\end{itemize}
These expressions provide the explicit forms of the Bardeen impact parameters $(\alpha, \beta)$ for the shadow boundary of a rotating wormhole, both in general and for a specific metric choice.

The complete shadow boundary is formed by the common envelope of the two curves mentioned above. The unstable circular orbits outside the throat correspond to one curve, while the throat critical orbits correspond to another curve. The shadow region is the common interior enclosed by both curves.

\section{Formation of the cuspy structure}
\label{sec:cusp}

In this section, we will explicitly study the shadow cast by a rotating traversable wormhole with redshift function shown in \eqref{eq:N}. The shadow will develop cuspy structures, and we will analyze the properties of the shadow.
For numerical convenience, we will set $G=c=1$ and $r_0=M=1$ throughout the section, thus we have $b=r_0=1$ and $a=cJ/(GM^2)=J$. 
In our analysis, we restrict consideration to observers situated on the equatorial plane with $\tob=\pi/2$. 
The qualitative features of the shadow remain unchanged for other observer inclinations; the general case can be readily recovered by tracking the parameter $\tob$.

The rotating traversable wormhole model being studied in this paper is characterized by two free parameters: the spin parameter $a$, which measures the angular momentum, and the redshift parameter $\lambda$, which controls the steepness of the redshift function as one approaches the throat from outside and thereby modulates the gravitational potential experienced by photons. 
Previous studies of rotating wormhole shadows have typically fixed the redshift function to a specific form (e.g., $\lambda = 0$) and focused on the influence of the spin parameter. 
Consequently, the possibility of a genuine morphological transition, from a smooth shadow boundary to one exhibiting a cusp, has never been reported for rotating wormholes. 
As we will demonstrate, allowing $\lambda$ to vary unveils a far richer phenomenology: the transition from a smooth to a cuspy shadow boundary becomes accessible, and the $(\lambda, a)$ parameter space hosts a strikingly rich phase structure comprising four distinct shadow morphologies.

\subsection{The development of the cusp with different $\lambda$}

\begin{figure}[H]
    \centering
    \subfloat[$\lambda=0.20$\label{21}]{\includegraphics[width=0.45\textwidth]{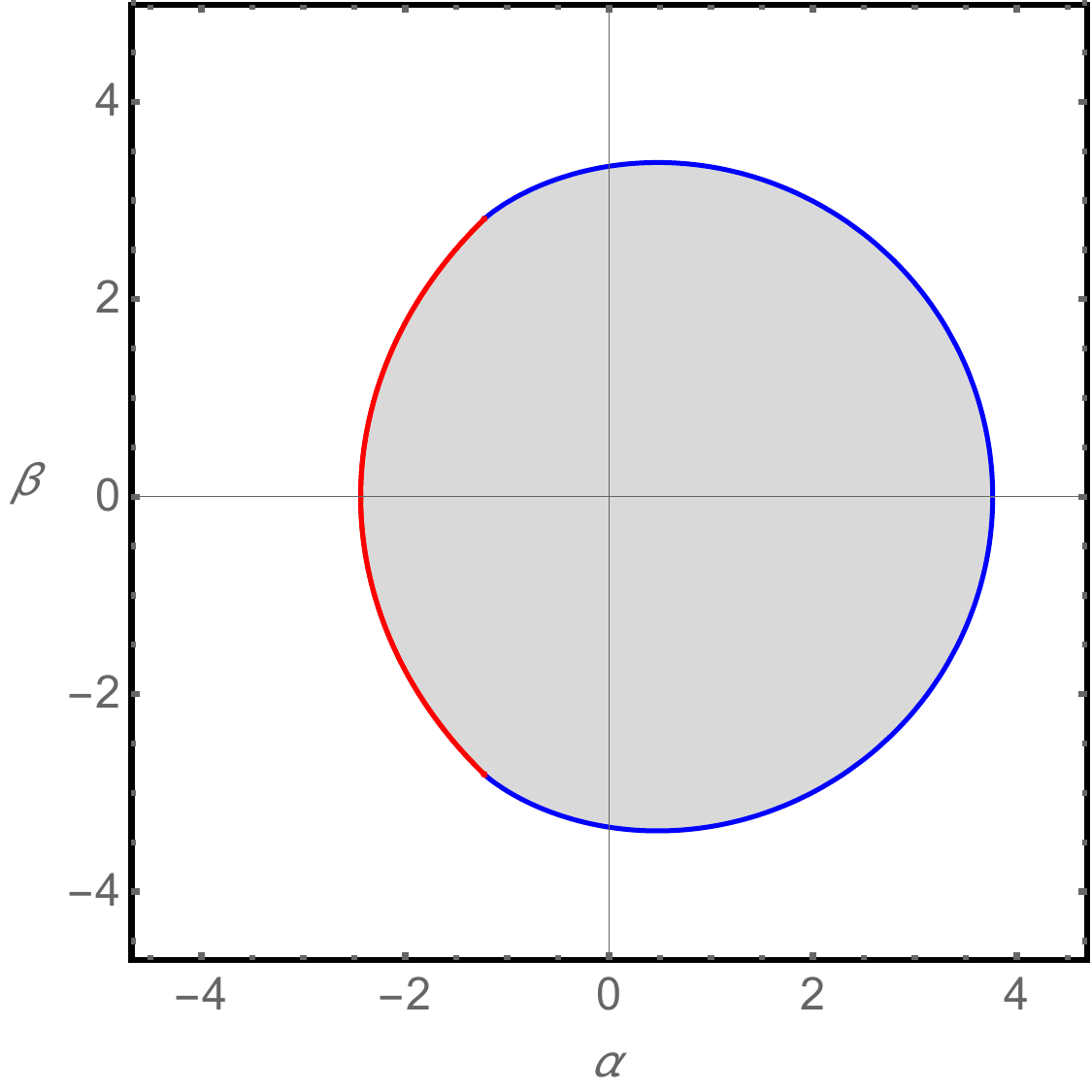}}\qquad
    \subfloat[$\lambda=\lambda_c=\frac{1}{4}(\sqrt{5}-1)$\label{22}]{\includegraphics[width=0.45\textwidth]{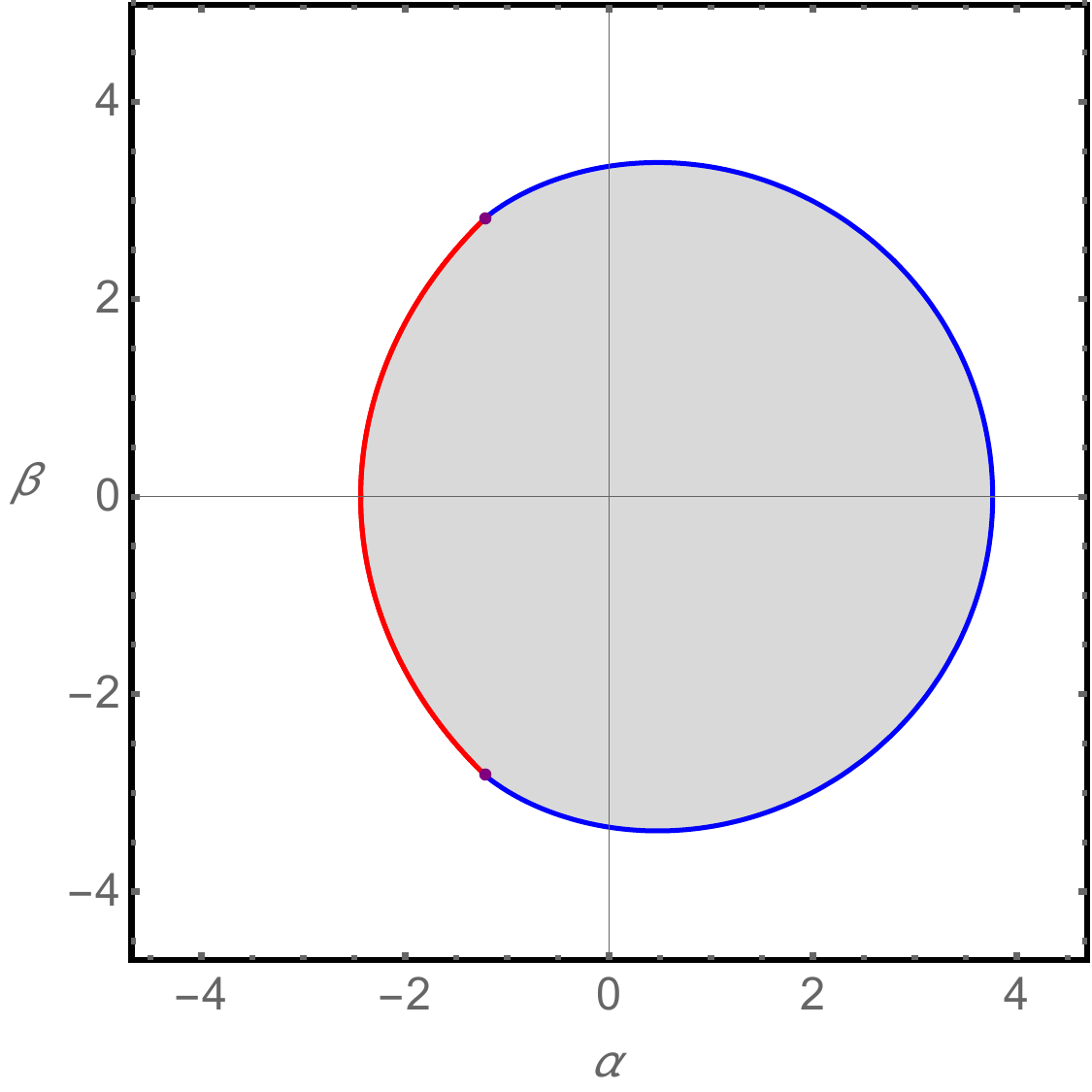}}\\
    \subfloat[$\lambda=0.80$\label{23}]{\includegraphics[width=0.45\textwidth]{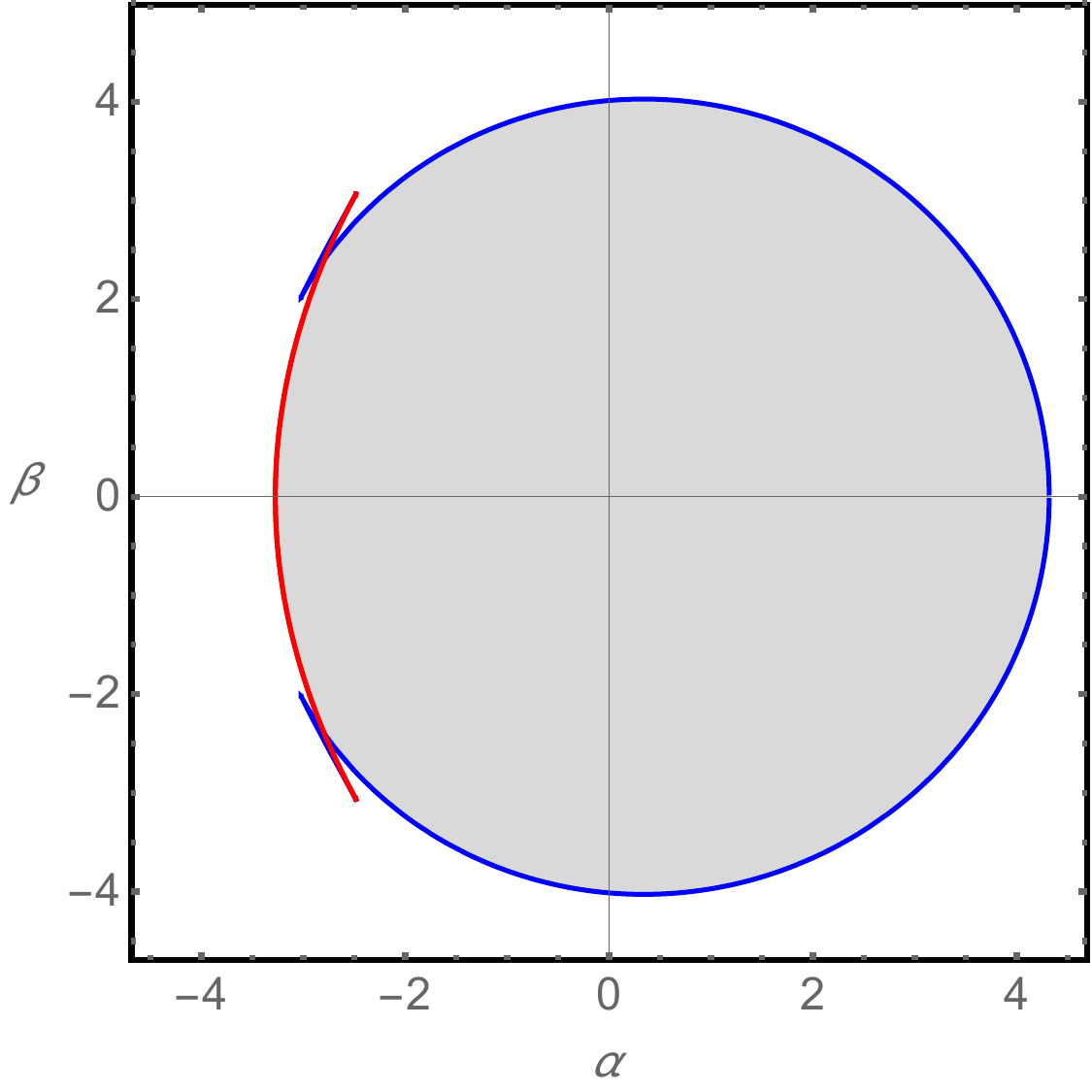}}\qquad
    \subfloat[$\lambda=\lambda_{\text{touch}}\approx 0.99$\label{24}]{\includegraphics[width=0.45\textwidth]{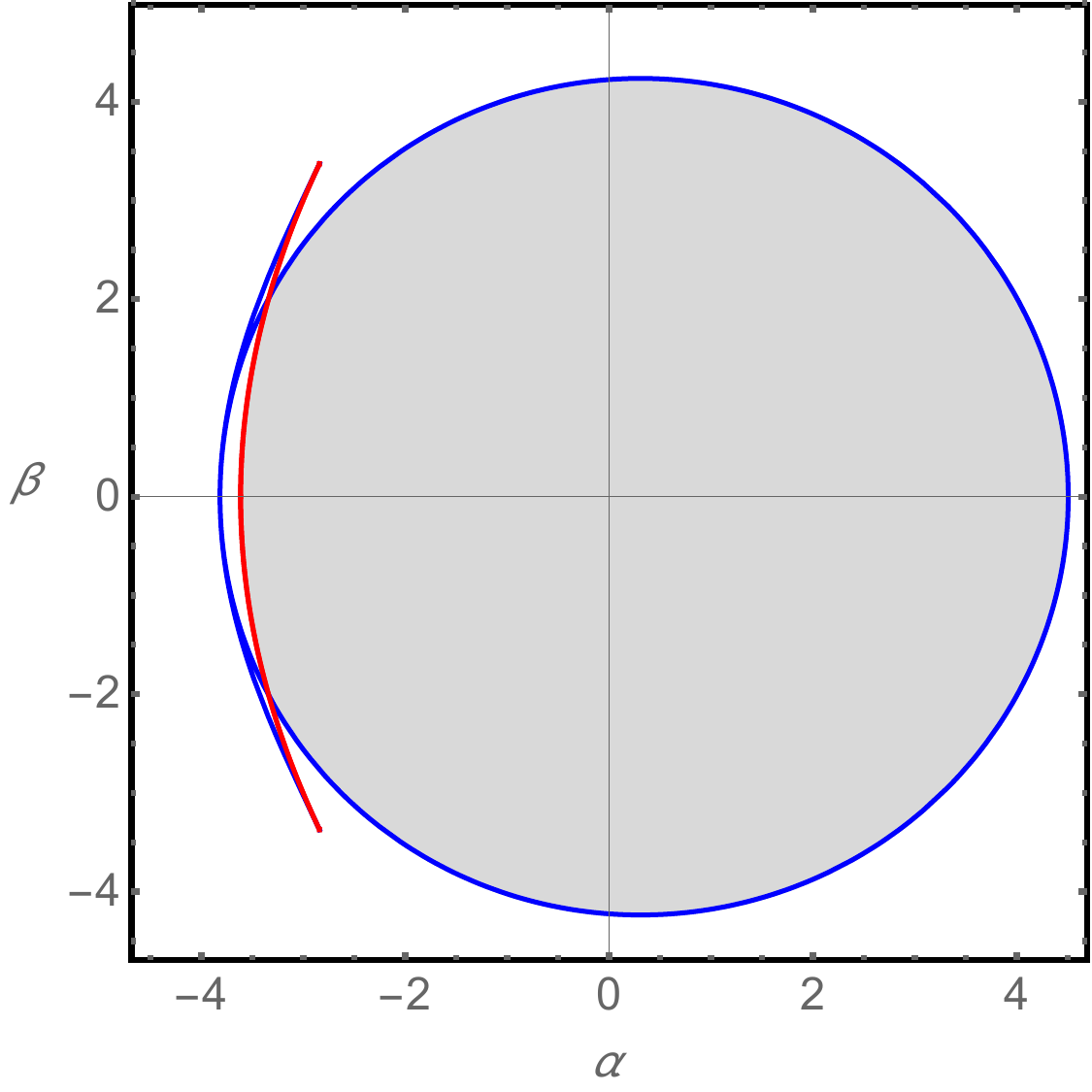}}\\
    \caption{The shadow of a rotating traversable wormhole with $a=0.07$. The shadow boundary is composed of two sets of orbits: the throat orbits (red curves) and the outer unstable circular orbits (blue curves). The real shadow is the shaded region enveloped by the two sets of curves.}
    \label{fig:smalla}
\end{figure}

Let us first take the spin parameter $a=0.07$, and study the change of the shadow as we gradually increase the redshift parameter $\lambda$. The shadow cast by a rotating wormhole is the common envelope of the unstable circular orbits outside the throat and the throat critical orbits. As illustrated in Fig. \ref{fig:smalla}, the size of the shadow grows with larger values of $\lambda$. There is a critical value of the redshift parameter 
\begin{equation}
	\lambda_c=\frac{1}{4} \left(\sqrt{5}-1\right)\approx 0.309\,.
\end{equation}
This critical value $\lambda_c$ is universal, independent of the specific choices for the spin parameter $a$, the redshift value $\lambda$, and the throat radius $r_0$. 

For $\lambda<\lambda_c$, for example $\lambda=0.2$, the lights from the unstable circular orbits and the wormhole throat form a smooth curve, as shown in Fig. \ref{21}. 
We can prove that the outer circular orbits and throat boundary approach each other with the same slope.
We define the slope of the shadow boundary, following \cite{Wei:2026zwu}, as
\begin{equation}
	\mathcal{F}=\frac{\dd \beta}{\dd \alpha}\,.
\end{equation}
So the slope of the outer circular orbits when approaching the throat orbits can be obtained from \eqref{eq:ab}
\begin{equation}\label{F1}
	\mathcal{F}_{\text{out}}=\frac{\dd \beta}{\dd \alpha}\Bigg{|}_{r_{\text{ph}}\to r_0}=\frac{6 e^{2 \lambda+2} a^2 \sin ^2\tob +\lambda r_0^4}{r_0^2 \sqrt{9 e^{2 \lambda+2} a^2 \sin ^2\tob-\lambda^2 r_0^4}}.
\end{equation}
From \eqref{eq:throat37}, the slope of the throat orbits when it approaches the outer circular orbits is
\begin{equation}\label{F2}
	\mathcal{F}_{\text{throat}}=\frac{6 e^{2 \lambda+2} a^2 \sin ^2\tob +\lambda r_0^4}{r_0^2 \sqrt{9 e^{2 \lambda+2} a^2 \sin ^2\tob-\lambda^2 r_0^4}}.
\end{equation}
Those two expressions shown in \eqref{F1} and \eqref{F2} are exactly the same. For $r_0=1$ and $\tob=\pi/2$, we have 
\begin{equation}
	\mathcal{F}_{\text{out}}=\mathcal{F}_{\text{throat}}=\frac{6 a^2 e^{2 \lambda+2}+\lambda}{\sqrt{9 a^2 e^{2 \lambda+2}-\lambda^2}}\,.\label{FF}
\end{equation}

\begin{figure}[H]
\centering
  \includegraphics[width=0.5\textwidth]{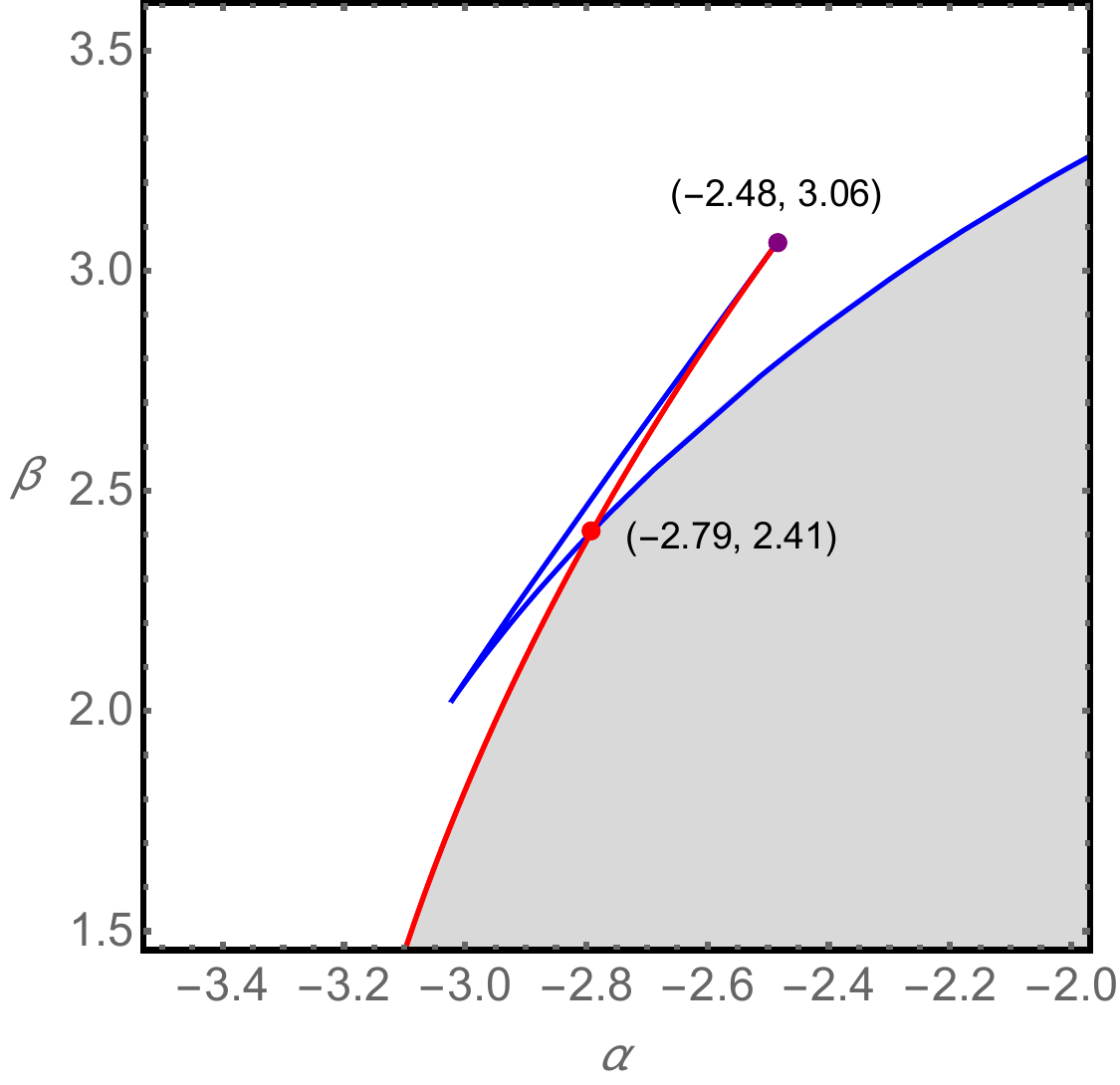}
  \caption{The cusp and swallowtail for the unstable circular orbits, with $a=0.07$ and $\lambda=0.8$. The outer unstable circular orbits (blue) and throat orbits (red) meet at the purple dot, with the same slope $\mathcal{F}_{\text{out}}=\mathcal{F}_{\text{throat}}$. The cusp is illustrated by the red dot.}\label{fig:cross}
\end{figure}

The critical point appears at $\lambda=\lambda_c$, as shown in Fig. \ref{22}. 
For values exceeding this threshold ($\lambda>\lambda_c$​), a cuspy structure develops and becomes increasingly distinct. We depict the shadow with $\lambda=0.8$ in Fig. \ref{23}, where the ``swallowtail'' behavior of the outer unstable circular orbits is explicitly demonstrated.
It is worth noting that the swallowtail structure itself, though present in the diagram, is not directly visible in the observational shadow.
There can be light coming from the triangle region enclosed by the red and blue curves shown in Fig. \ref{fig:cross}. 
What we can observe is a cusp of the shadow located at $(\alpha,\beta)=(-2.79,2.41)$, for the case with $a=0.07$ and $\lambda=0.8$. 
The outer unstable circular orbits and throat orbits meet at the purple dot $(\alpha,\beta)=(-2.48,3.06)$, as demonstrated in Fig. \ref{fig:cross}. Additionally, the slopes of the two curves are the same, as discussed in equation \eqref{FF}.
The cusp where the two curves intersect is illustrated by the red dot in the figure.
The slopes of the two curves are different at the cusp point, and we have a slope discontinuity, hence the term ``cusp''.

\begin{figure}[H]
    \centering
    \subfloat[$\lambda=\lambda_{\text{drown}}\approx 1.38$\label{41}]{\includegraphics[width=0.45\textwidth]{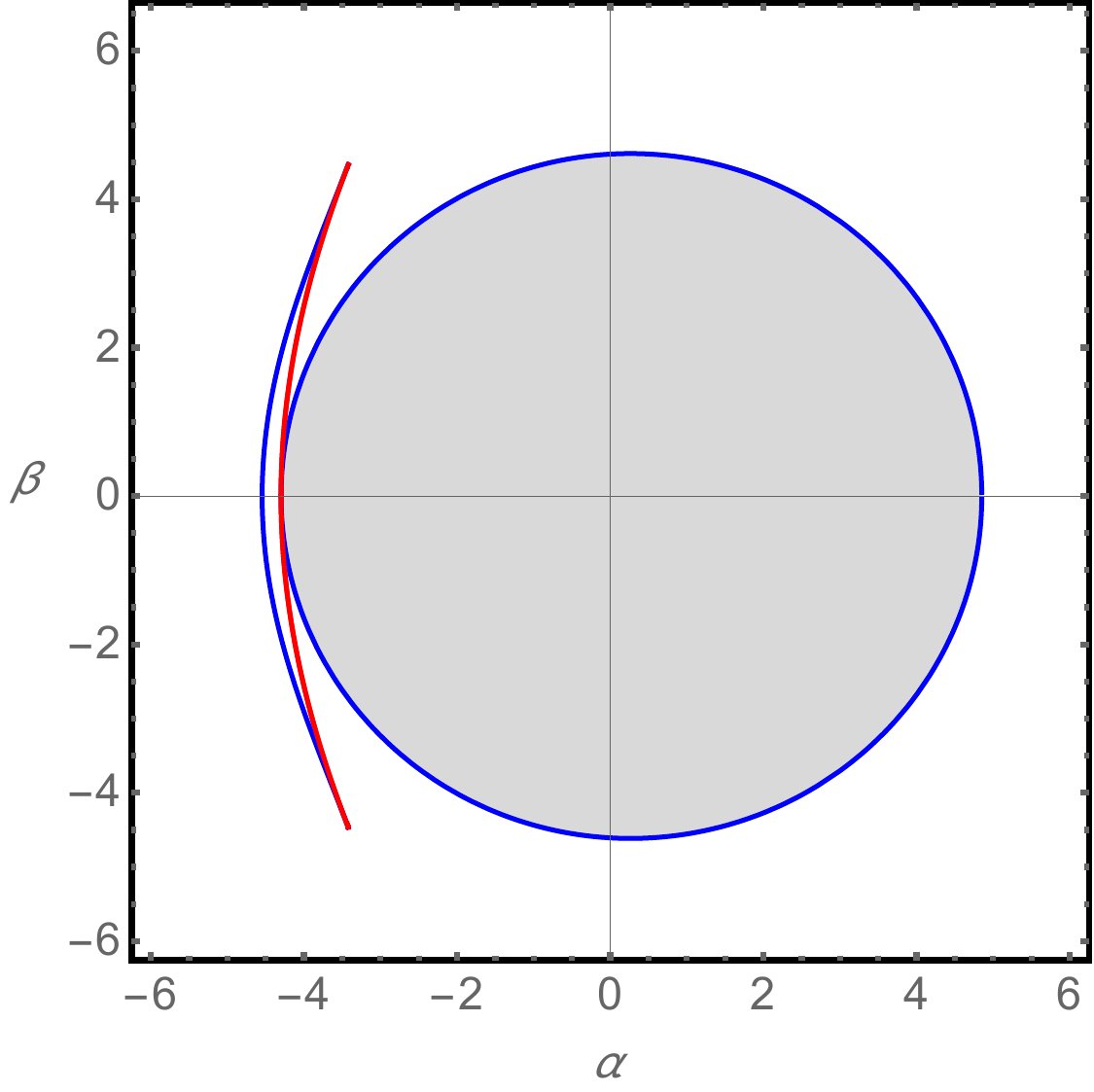}}\qquad
    \subfloat[$\lambda=1.70$]{\includegraphics[width=0.45\textwidth\label{42}]{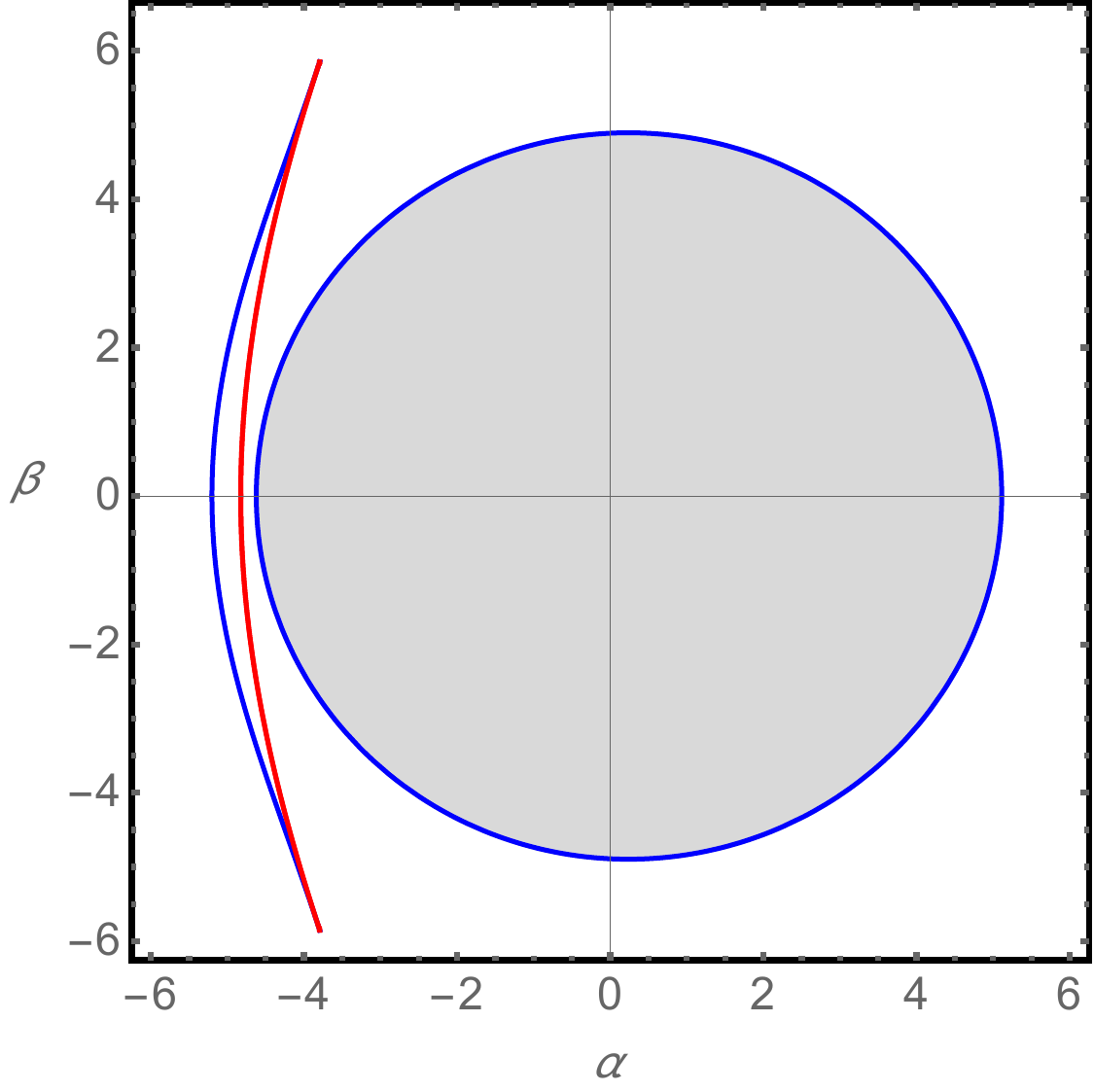}}\\
    \subfloat[$\lambda\approx 5.41$\label{43}]{\includegraphics[width=0.45\textwidth]{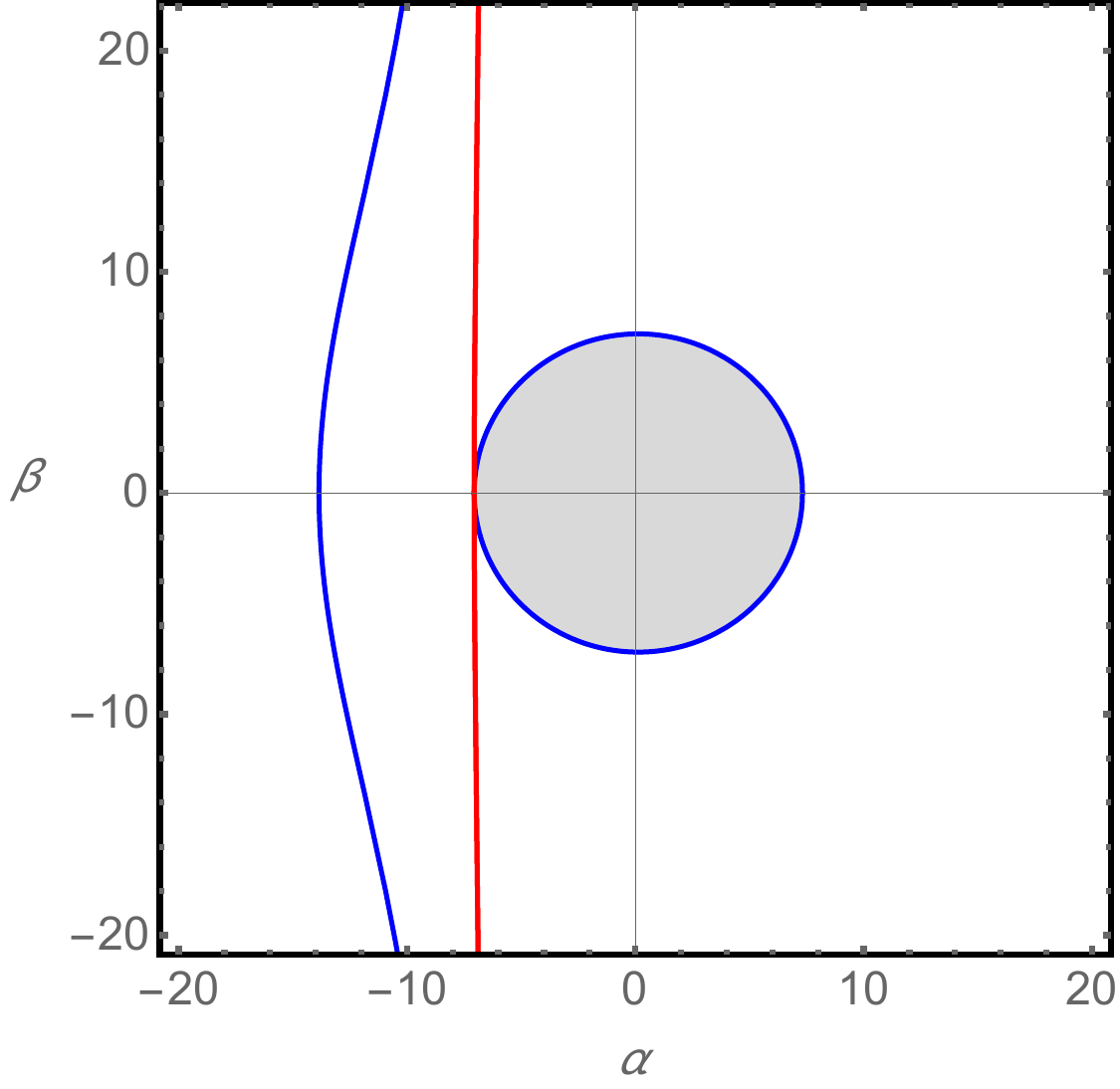}}\qquad
    \subfloat[$\lambda=6.00$\label{44}]{\includegraphics[width=0.45\textwidth]{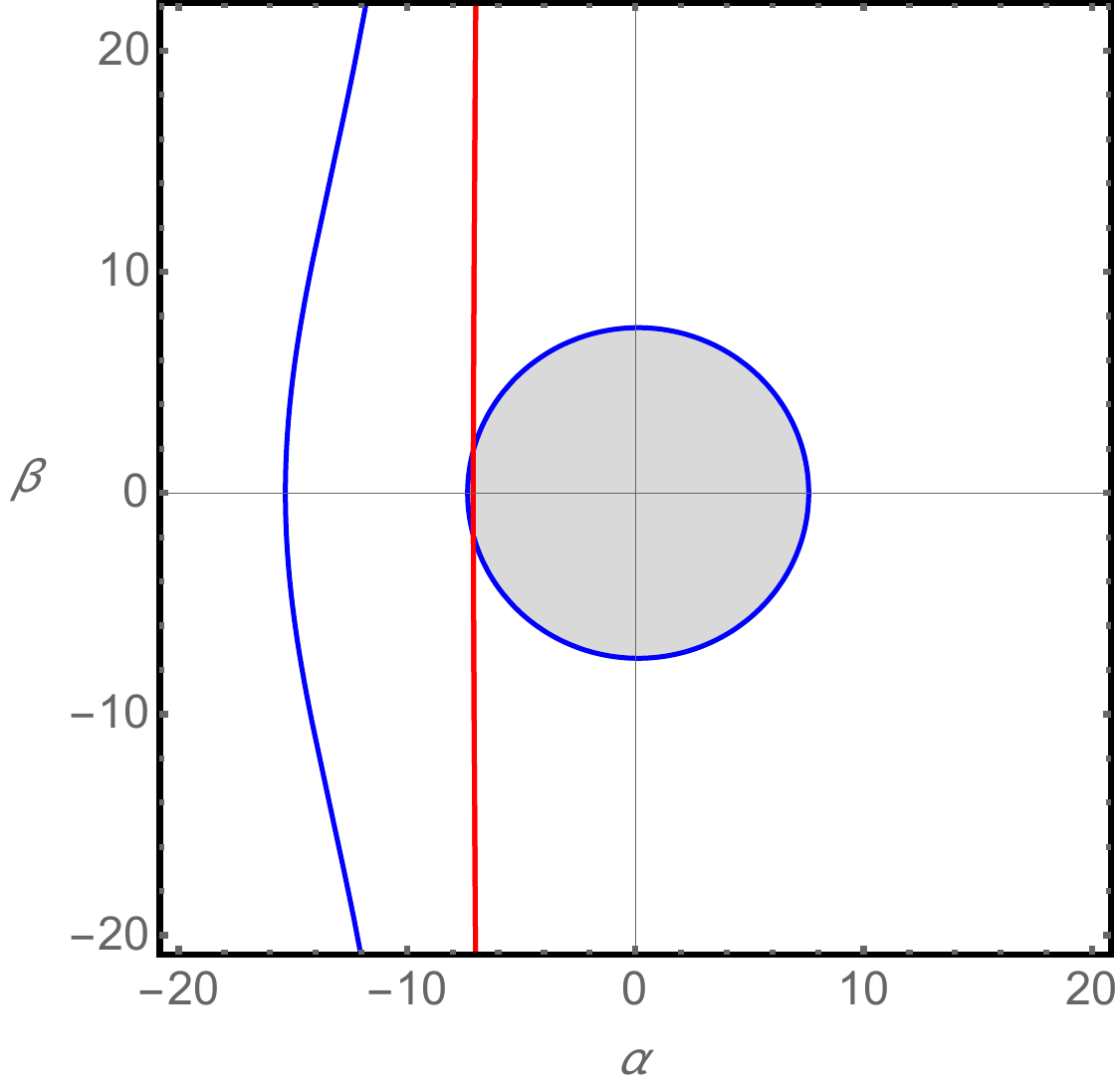}}\\
    \caption{The rotating traversable shadow with $\lambda$ (with $a=0.07$). 
    For $\lambda\geq \lambda_{\text{drown}}$, with increasing $\lambda$, the curve of the wormhole throat orbits first detaches from the shadow, and then becomes tangent with the shadow at $\lambda=5.41$.
    For even larger $\lambda$, the shadow boundary comprises the throat orbits and the outer unstable circular orbits.}
    \label{fig:4}
\end{figure}

For larger redshift parameter $\lambda$, the ``ears'' of the swallowtail start to touch each other at a critical value $\lambda_{\text{touch}}\approx 0.99$, as shown in Fig. \ref{24}. 
Although this is not an observable behavior, the phenomenon is interesting to study. 
At $\lambda_{\text{touch}}$, the blue curve representing the outer unstable circular orbits, seals off into a quasi-circular shape.
As displayed in Fig. \ref{24}, as we raise the photon orbit radius $r_{\text{ph}}$ from $r_0=1$, $\beta$ decreases to zero and then starts to increase afterward. 
For $\lambda>\lambda_{\text{touch}}$, the blue curve breaks into two disjoint segments: a quasi-circular curve and a curve connecting the throat orbits, which is plotted in Fig. \ref{63}.

As we keep increasing $\lambda$, the throat orbits represented by the red curve would decouple from the blue quasi-circular curve, such that the shadow is solely enveloped by the outer unstable orbits, as displayed in Fig. \ref{41}.
The wormhole throat recedes from view and does not play any role in the shadow. We can call this situation ``throat drowning'' and denote the critical value of $\lambda$ as $\lambda_{\text{drown}}$.
For even larger $\lambda>\lambda_{\text{drown}}$, for instance $\lambda=1.7$ shown in Fig. \ref{42}, the triangle region enclosed by the blue and red curves separates from the wormhole shadow.
One may expect the region related to the wormhole throat to be separated forever.
This expectation, however, is not borne out, and we can see from Fig. \ref{43} that the wormhole throat orbits reconnect with the shadow again at $\lambda=5.41$. For $\lambda>5.41$, the wormhole contributes to the shadow boundary again, and the shadow becomes the common envelope of the two curves again, for large $\lambda$. 
Note that for large $\lambda$, the throat orbits behave like a vertical line with fixed $\alpha$ as shown in Fig. \ref{44}. This tendency becomes increasingly pronounced as $\lambda$ grows further. The throat-drowning situation, however, is not common for all rotating wormholes. As will be demonstrated subsequently, the throat orbits submerge only for sufficiently small spin parameters.
For spin parameter $a>0.08$, the wormhole throat would invariably contribute to the shadow boundary.

\subsection{Phase diagram of the shadow boundary}

After a thorough discussion of the shadow morphology with a fixed spin parameter, one may be prompted to wonder if there is a phase diagram describing all the different phases demonstrated in the previous subsection.
It was shown in \cite{Wei:2026zwu} that the formation of the cusps for a rotating black hole corresponds to a topological phase transition.  Thus, we use the term ``phase'' to represent different behaviors of the shadow boundary.
We illustrate different phases of the wormhole shadow and discuss some essential properties here.

\begin{figure}[H]
    \centering
    \includegraphics[width=0.7\textwidth]{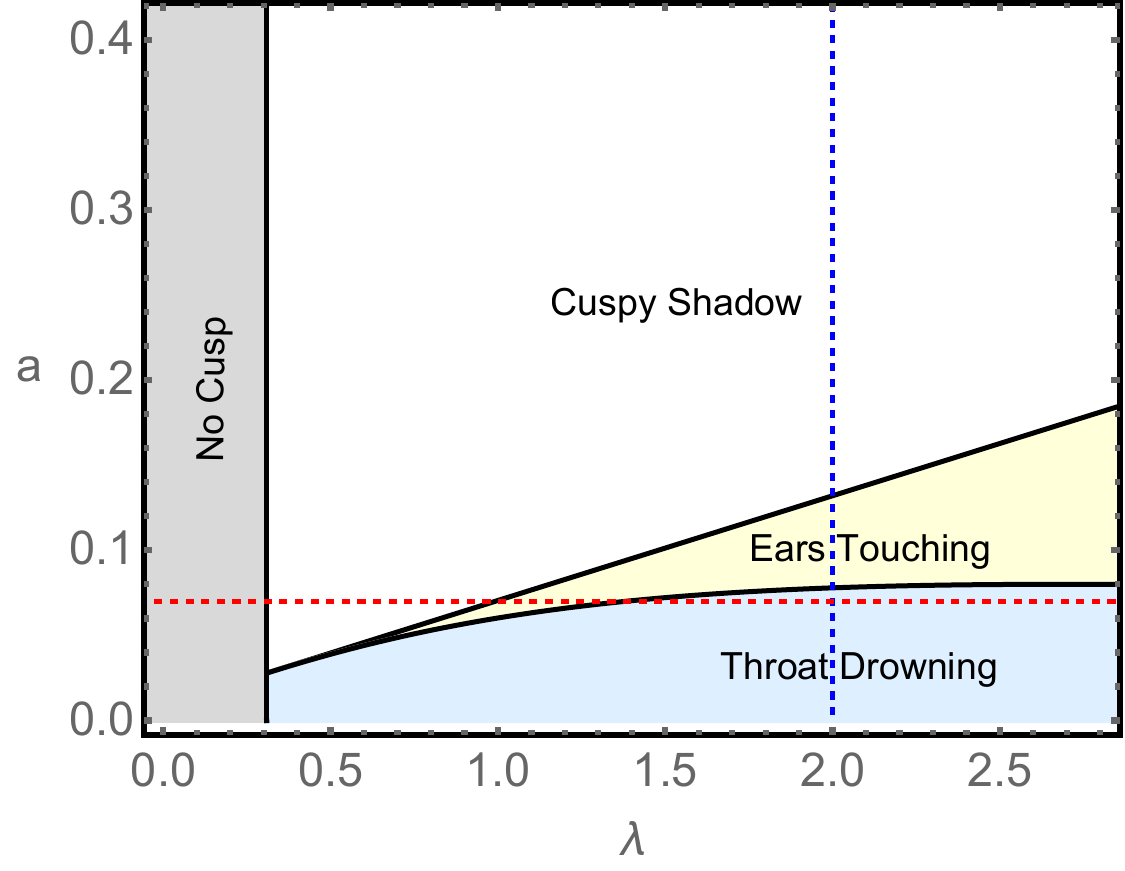}
    \caption{Phases of the shadow boundary illustrated with different $a$ and $\lambda$.
    For $\lambda<\lambda_c$, the shadow boundary is smooth. For larger $\lambda$, there are three different phases: cuspy shadow, ears touching, and throat drowning. The blue dotted line represents $\lambda=2$, and the red dotted line is $a=0.07$. }
    \label{fig:PD}
\end{figure}

\begin{figure}[H]
    \centering
    \subfloat[$a= 0.500$\label{61}]{\includegraphics[width=0.45\textwidth]{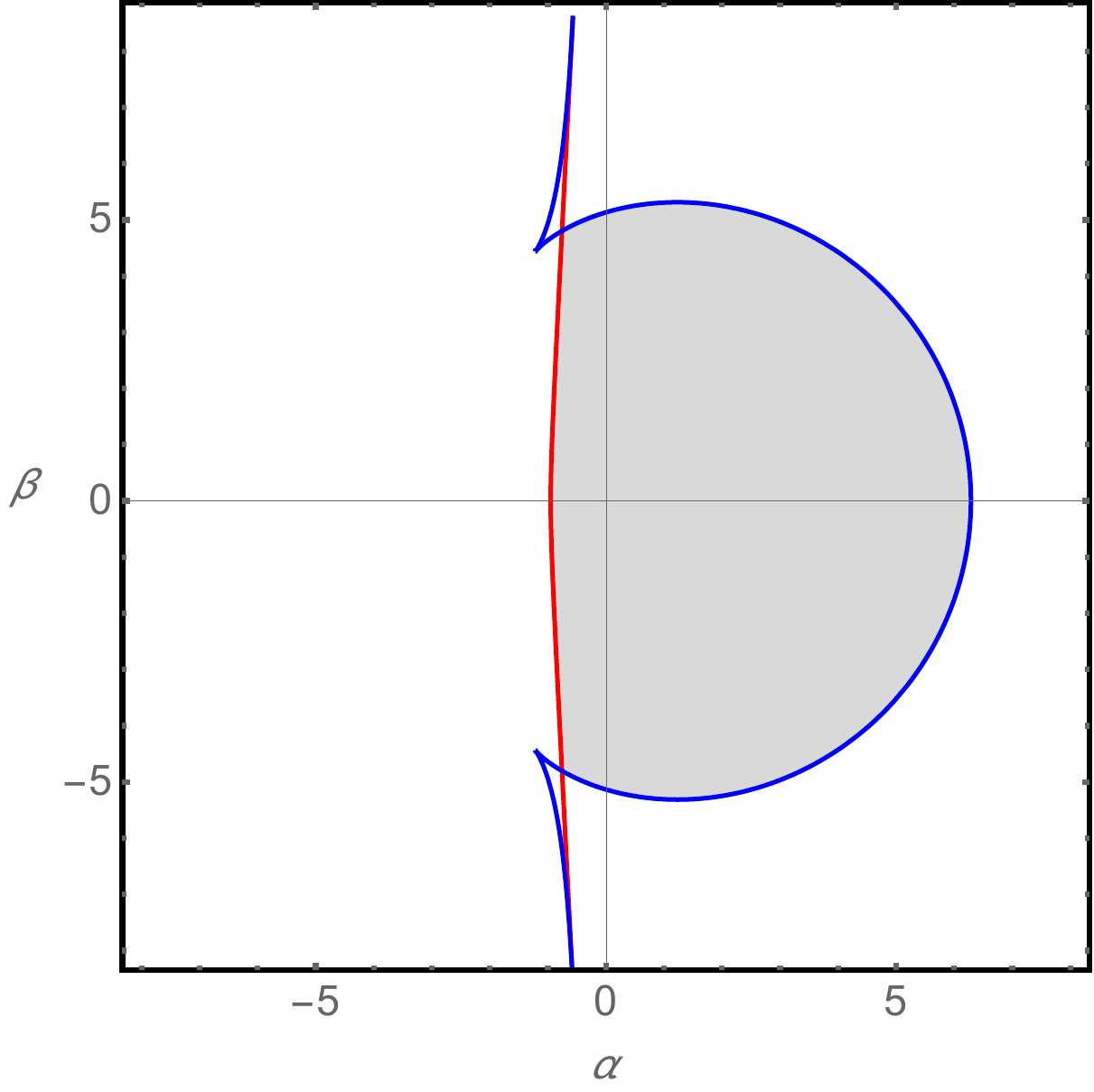}}~~
    \subfloat[$a\approx 0.132$\label{62}]{\includegraphics[width=0.45\textwidth]{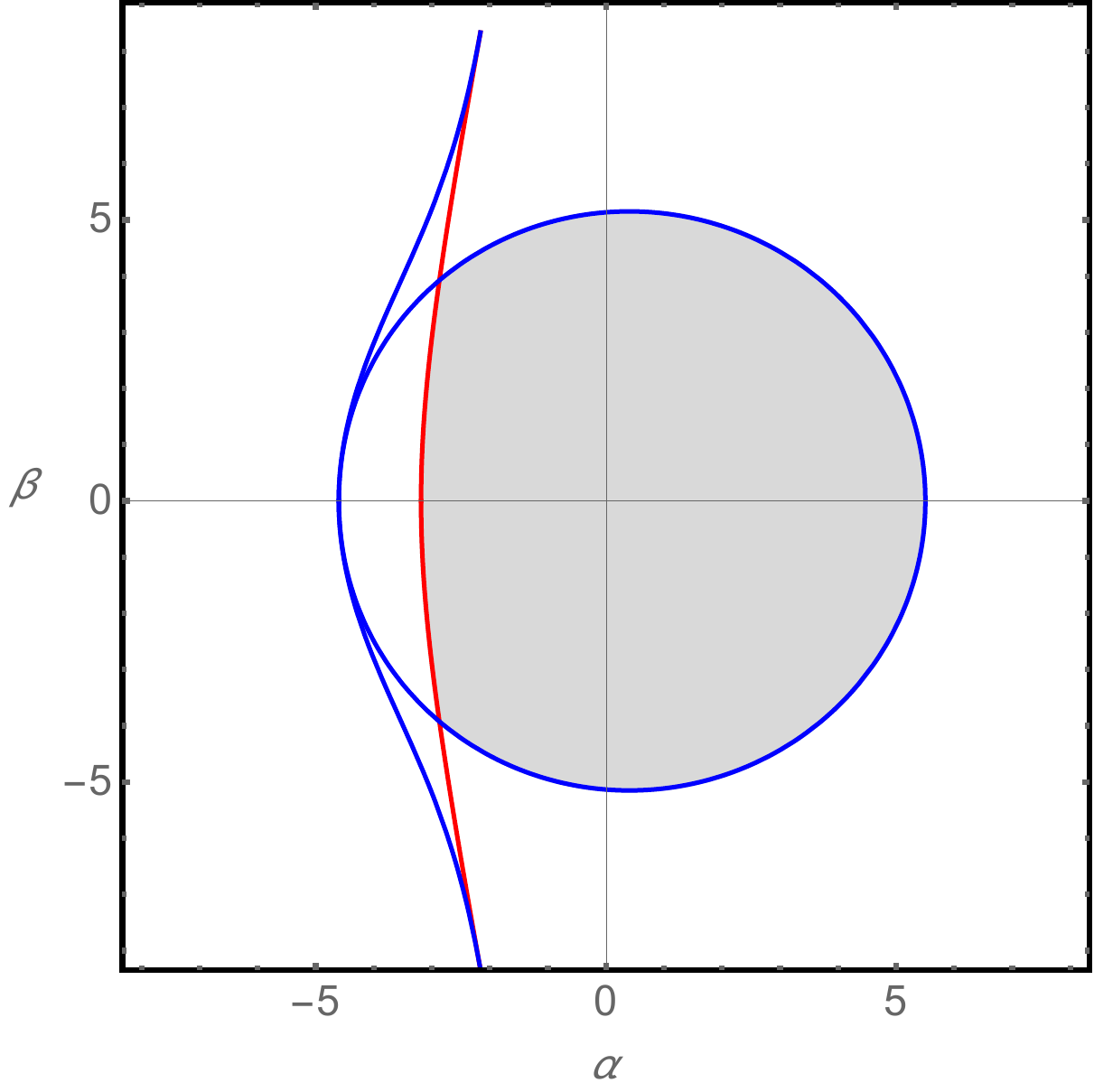}}\\
    \subfloat[$a=0.100$]{\includegraphics[width=0.45\textwidth\label{63}]{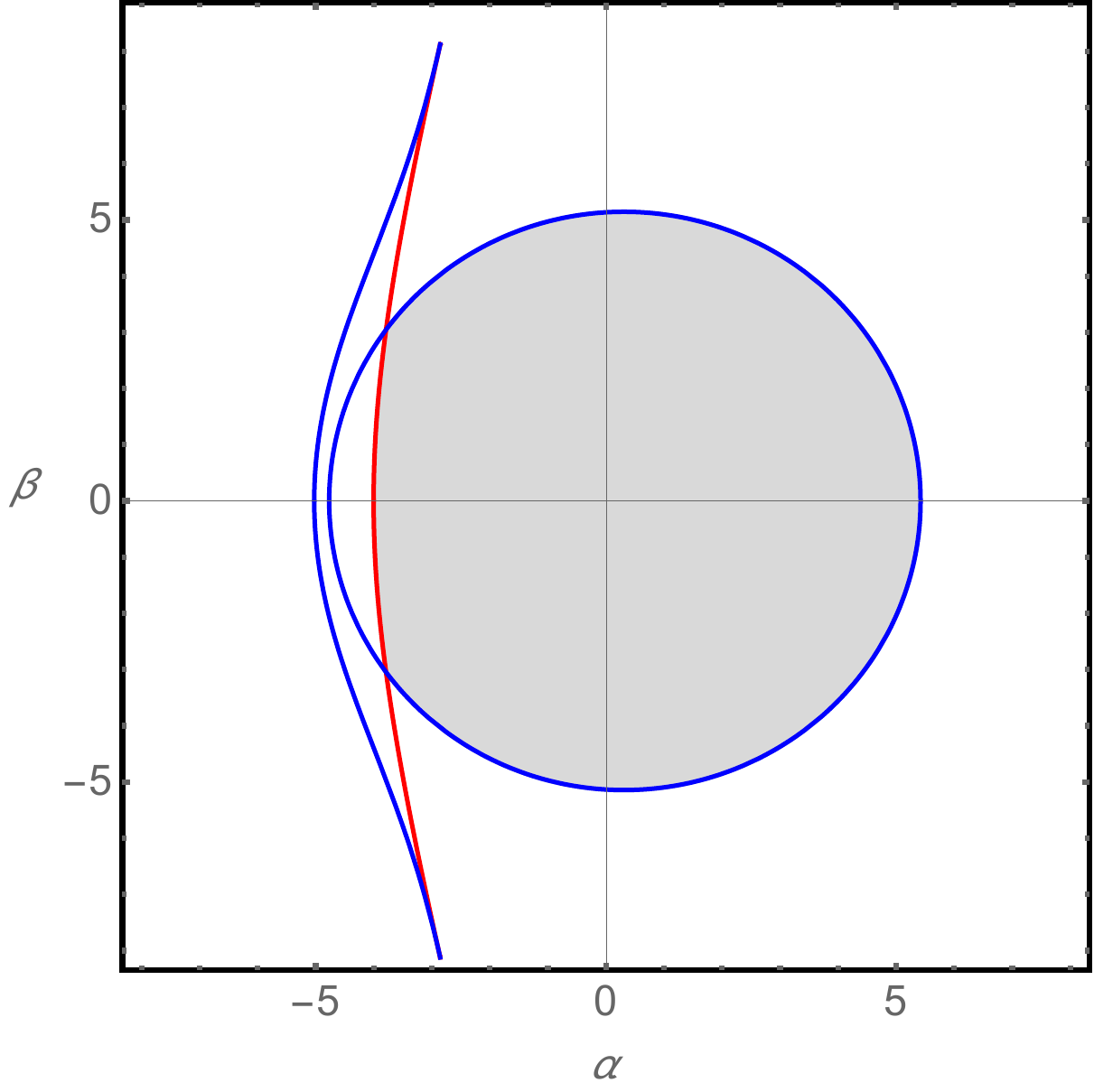}}~~
    \subfloat[$a\approx 0.078$]{\includegraphics[width=0.45\textwidth\label{64}]{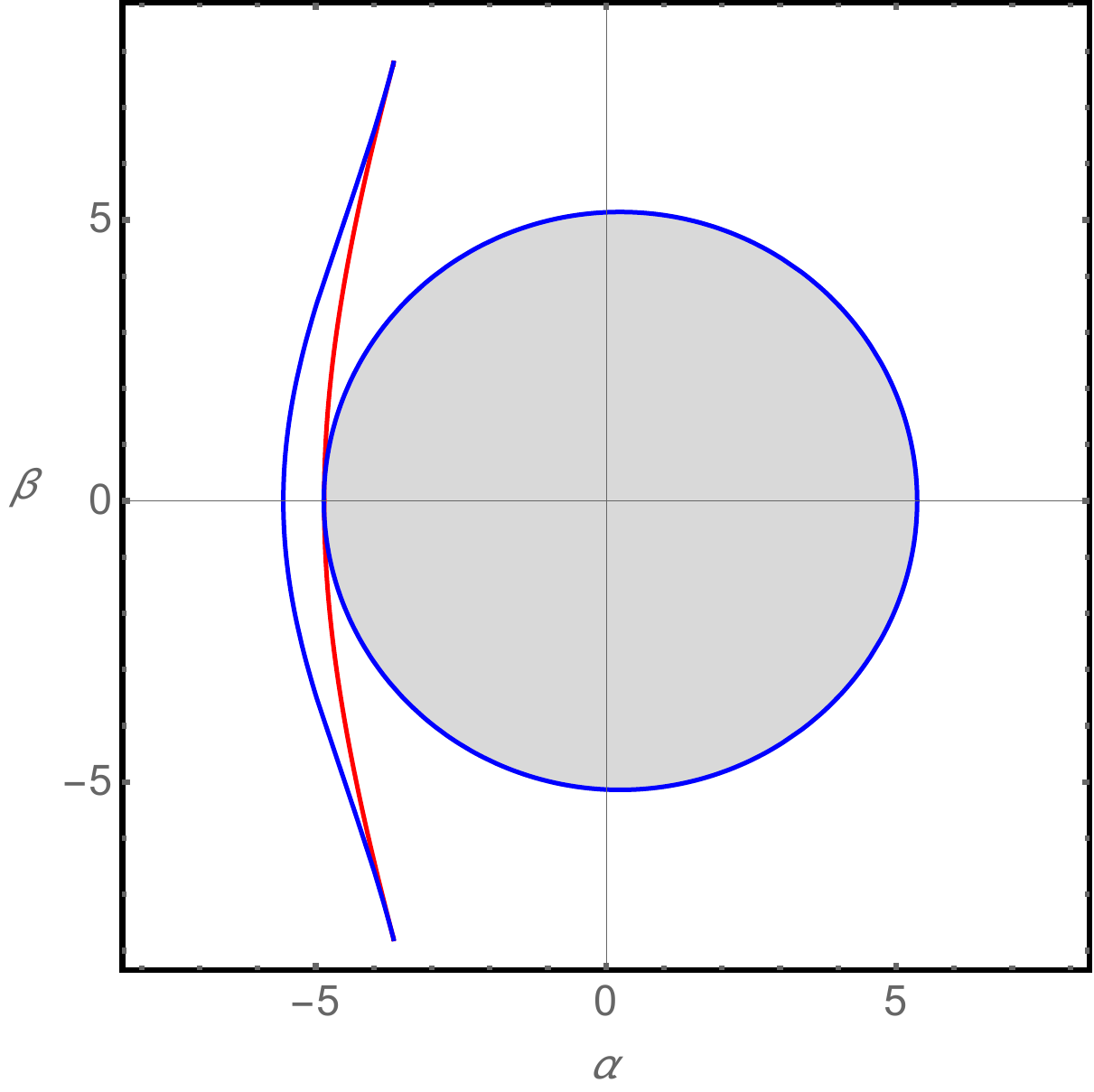}}
    \caption{The wormhole shadow boundary for fixed $\lambda=2$. The figure corresponds to the vertical dotted line in Fig. \ref{fig:PD}. }
    \label{fig:lambda}
\end{figure}

The diagram representing different phases of the wormhole shadow boundary is illustrated in Fig. \ref{fig:PD}. As can be seen from the figure, for $\lambda<\lambda_c=\frac{1}{4}(\sqrt{5}-1)$, no matter what the spin parameter $a$ is, the shadow boundary is smooth, and the shadow exhibits no cusp structure. This is consistent with previous studies where the model with $N=\exp(-r_0/r)$ is considered \cite{Shaikh:2018kfv,Gyulchev:2018fmd}.
As $\lambda$ continues to increase, the cusp starts to form at the critical value $\lambda_c$.
For larger $\lambda$, there are three different phases, which are cuspy shadow, ears touching, and throat drowning, respectively.
The blue dotted line in Fig. \ref{fig:PD} illustrates different phases with fixed $\lambda=2$, and the corresponding morphology of the shadow boundary is displayed in Fig. \ref{fig:lambda}. 
The red dotted line represents the situations discussed in the previous subsection with $a=0.07$.
As discussed before, for fixed $a=0.07$, the cusps start to appear at $\lambda=\lambda_c=\frac{1}{4}(\sqrt{5}-1)$ as we increase $\lambda$. Keep increasing $\lambda$, the ears of the shadow boundary start to get connected at $\lambda=\lambda_{\text{touch}}\approx 0.99$. For $1.38<\lambda<5.54$, the throat can not be seen by the shadow observation, and we are in the throat-drowning region. The shadow is determined solely by the outer unstable orbits and is smooth in this region. 
For $\lambda>5.54$, the wormhole throat orbits get reconnected to the shadow, and the shadow becomes enclosed by the two families of orbits with cusps, as shown in Fig. \ref{44}.

\begin{figure}[H]
\centering
\includegraphics[width=0.55\textwidth\label{distouch}]{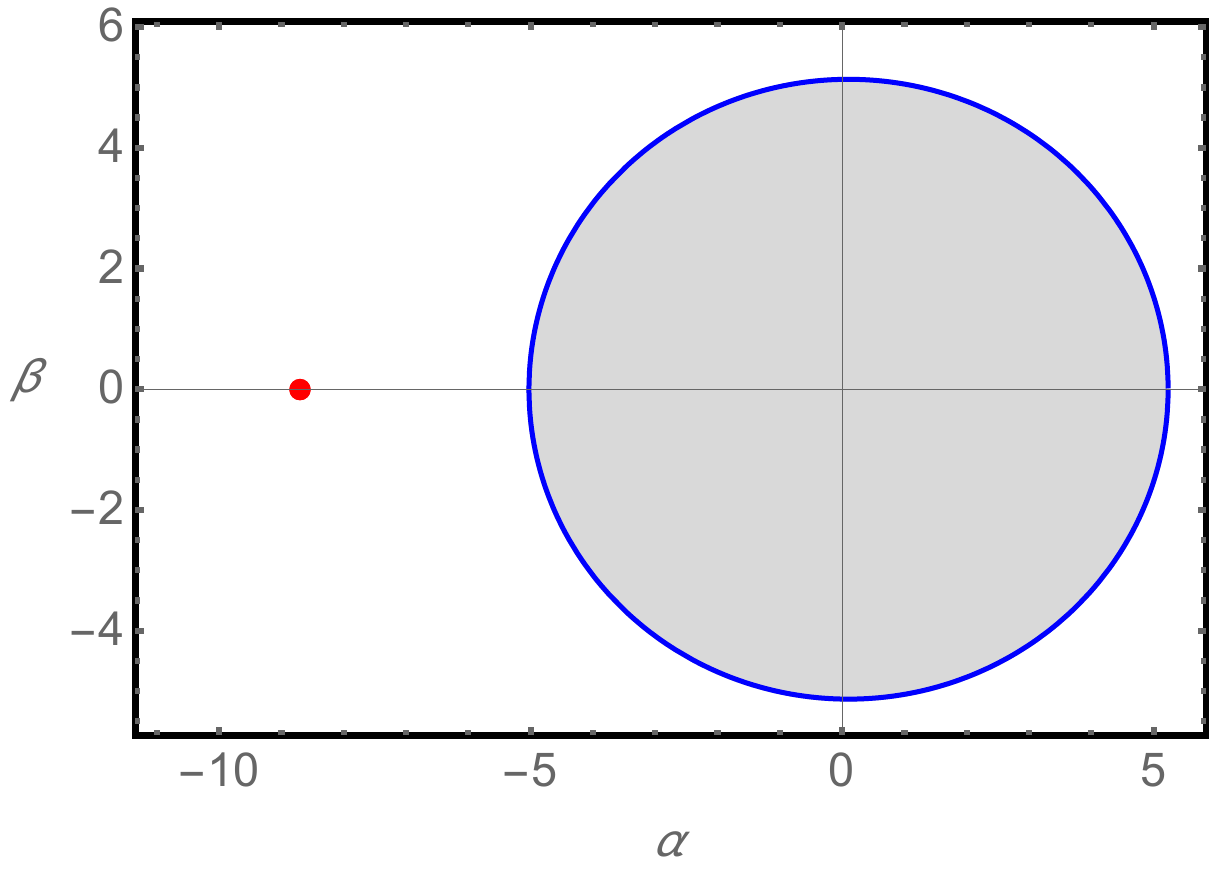}
  \caption{The wormhole shadow with $\lambda=2$ and $a\approx 0.033$ . For this critical value of spin parameter, the wormhole throat orbits shrink to a single point. The throat orbits completely disappear for even smaller $a$. }
  \label{fig:disappear}
\end{figure}

We now discuss these three different phases illustrated by the blue dotted line in Fig. \ref{fig:PD} as we decrease the spin parameter:
 \begin{itemize}
  \item For a rotating wormhole with a large spin parameter, there are cuspy structures for the wormhole shadow, and the outer unstable circular orbits exhibit swallowtail behavior as shown in Fig. \ref{61} and Fig. \ref{fig:cross}. In this case, the shadow is the shaded region enclosed by the wormhole throat orbits and the outer unstable orbits.
  \item As we decrease the spin parameter, the ears of the shadow boundary start to touch each other, as illustrated in Fig. \ref{62}. The parameter region when the two ears of the boundary shadow are connected is called ``ears touching'' phase, which is shown by the yellow region in Fig. \ref{fig:PD}. It is worth emphasizing that the critical line representing the ears starting to touch each other is a straight line in the phase diagram, implying a linear dependence between parameters $a$ and $\lambda$. The slope of the ears-touching line is approximately 0.06.
  \item In the ears touching region, the shadow boundary is a composite contour that consists of two distinct blue and red segments, as shown in Fig. \ref{63}. However, as we continue decreasing the spin parameter, after a threshold of $a$, the wormhole throat orbits recede from the shadow. For the $\lambda=2$ case, the critical value of the spin parameter is $a\approx 0.078$, which is displayed in Fig. \ref{64}. We call the region where the throat orbits detach from the shadow the throat-drowning region, which is represented by the blue area in Fig. \ref{fig:PD}.
  The line between the ears-touching and the throat-drowning regions demonstrates the situation when the red curve starts to detach from the shadow shown in Fig. \ref{64}. There would not be a throat-drowning situation for high-spinning wormholes, regardless of the value of $\lambda$. The line between the yellow and blue regions has a maximum at $(\lambda,a)\approx (2.68, 0.08)$.
  \item In the throat-drowning region, the wormhole throat orbits never contribute to the shadow. Moreover, for very small $a$, the wormhole throat orbits may shrink to a single point as displayed in Fig. \ref{fig:disappear}. For the case $\lambda=2$, the critical value for wormhole throat orbits shrinking to a point is $a\approx 0.033$.
  	It is worth emphasizing the difference between situations illustrated in Fig. \ref{64} and in Fig. \ref{fig:disappear}. 
  	Both cases are in the throat-drowning region, meaning the wormhole throat orbits would not contribute to the wormhole shadow.
  	For the case with $0.033\leq a\leq0.078$, the wormhole throat orbits detach from the shadow, thus they can not be observed from the shadow. However, the throat orbits are still there, and one may receive light from the wormhole throat. 
  	For $a<0.033$, the wormhole throat orbits completely disappear, and no light signal from the throat can be received.
\end{itemize}

\begin{figure}[h]
  \centering
   \centering
    \subfloat[$a=0.95$ and $\lambda\approx 15.33$]{\includegraphics[width=0.45\textwidth]{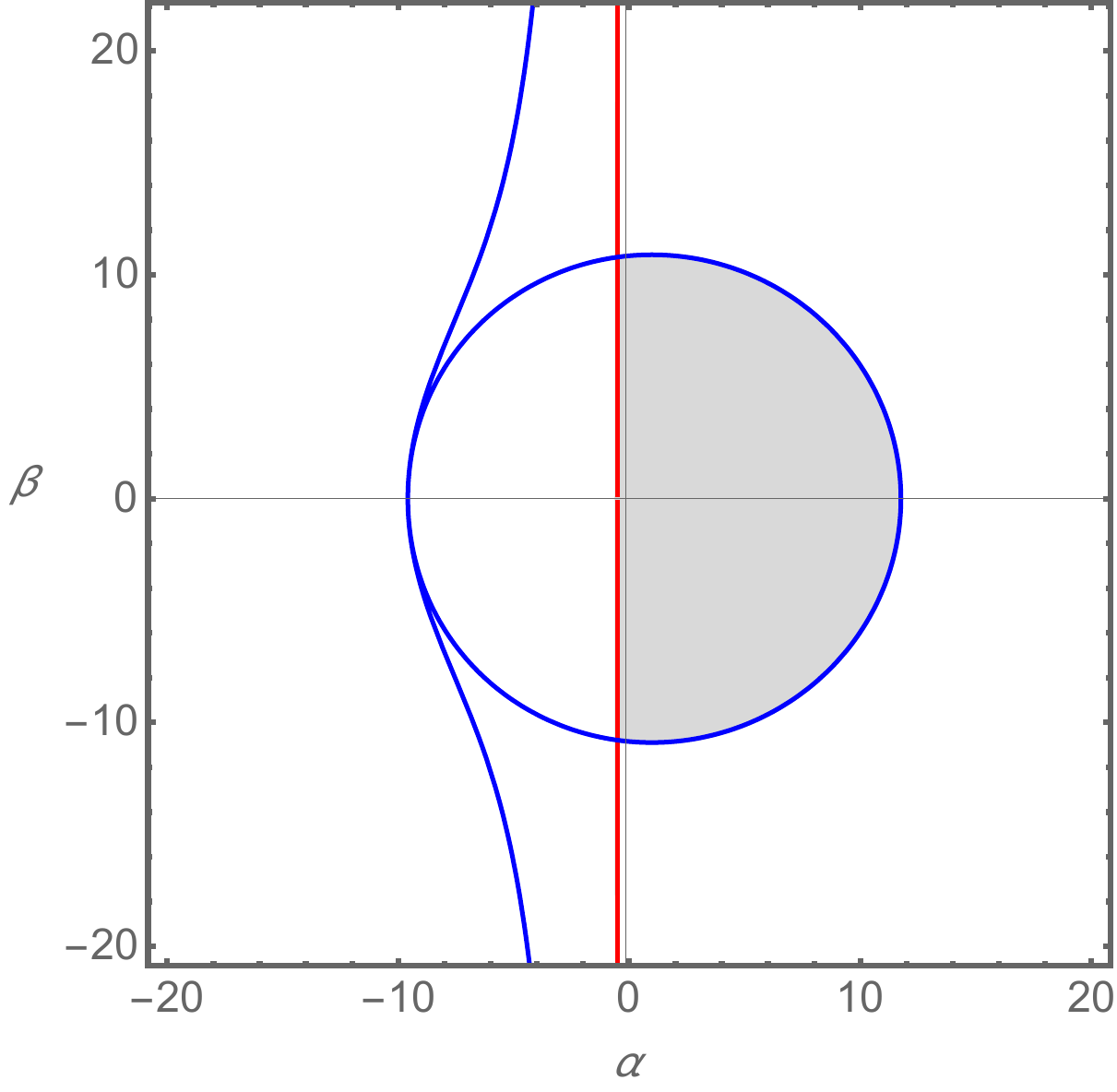}}~~~~
    \subfloat[$a=0.05$ and $\lambda\approx 0.67$]{\includegraphics[width=0.44\textwidth]{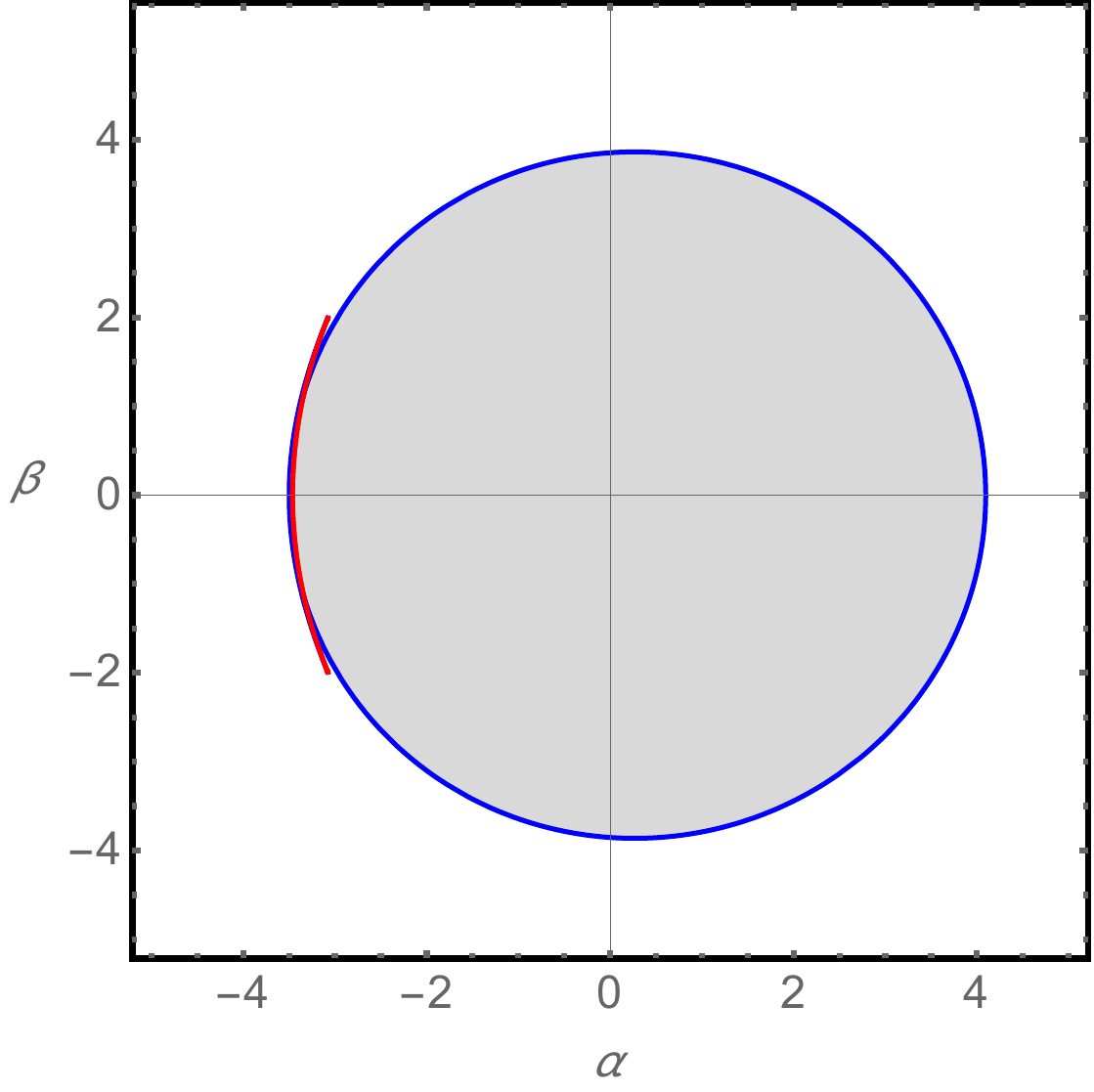}}
  \caption{The shadows with ears-touching feature for different values of $a$ and $\lambda$. }\label{fig:compare}
\end{figure}

It is important to describe the overall morphology and global deformation of the shadow as the spin parameter $a$ and the redshift parameter $\lambda$ vary.
The transition from a smooth to a cuspy shadow boundary, and indeed the very formation of the cusp itself, only emerges upon varying the redshift parameter $\lambda$.
As shown in Fig. \ref{fig:smalla}, the size of the shadow grows as we increase $\lambda$. 
The spin parameter does not have much influence on the size of the shadow, as can be seen from Fig. \ref{fig:lambda}. 
The primary role of the spin parameter is to modulate the horizontal asymmetry of the shadow about the vertical axis.
In Fig. \ref{fig:lambda}, for small $a$ like figures \ref{63} and \ref{64}, the shadow exhibits near-perfect left-right symmetry about the y-axis. While for relatively larger $a$, the wormhole shadow exhibits a pronounced displacement in the positive $\alpha$ direction, see Fig. \ref{61} for example.
We have discussed that for large $\lambda$, the wormhole throat orbits behave like a vertical line. Moreover, the size of the swallowtail boundary also increases for large values of $\lambda$. 
This contrast is further illustrated in Fig. \ref{fig:compare}, where two representative cases are shown.
For $a=0.95$ and $\lambda\approx 15.33$, the size of the swallowtail is huge, and the shadow exhibits a semicircular disk morphology.
For parameters $a=0.05$ and $\lambda\approx 0.67$, the swallowtail can barely be seen, and the shadow is symmetrically distributed.
The close relation between the shadow morphology and the two parameters $\lambda$ and $a$, provides a clear observational signature that may help distinguish the two parameters from future high-resolution images.

For small $a<0.08$, we have seen a re-entrant behavior of the throat orbits, detachment followed by reattachment, to the shadow. 
 Notably, for relatively large $\lambda$, the throat orbits reconnect with the shadow and behave almost as a vertical line in the $(\alpha, \beta)$ plane, while the outer orbits continues to expand. 
This phenomenon can be understood as a non-trivial competition between the redshift effect (which enlarges the shadow) and the frame-dragging effect (which distorts its shape).

\section{Conclusion and discussion}
\label{sec:con}

In this work, we have conducted a systematic investigation into the shadow cast by a class of rotating traversable wormholes, with particular emphasis on the formation of cuspy structures along the shadow boundary. 
We specialized in a concrete family of wormhole solutions with a redshift function of the form 
\begin{equation}
	N = \exp\left[-\frac{r_0}{r} - \lambda \left(\frac{r_0}{r} \right)^2\right]\,,
\end{equation}
which captures the essential physics and allows for an explicit analytical treatment. 
Unlike previous studies that fixed the redshift function and focused solely on the spin parameter, we consider $\lambda$ as a crucial degree of freedom and demonstrate that it governs genuine morphological transitions.
The shadow boundary is determined by the common envelope of two distinct families of critical curves: unstable circular photon orbits residing outside the throat, and critical orbits located exactly at the throat itself. 
By solving the null geodesic equations via the Hamilton-Jacobi formalism and mapping the impact parameters to Bardeen's celestial coordinates $(\alpha, \beta)$, we have obtained explicit parametric expressions for both families of orbits and analyzed their interplay as functions of the redshift parameter $\lambda$ and the spin parameter $a$.

  By performing a two-parameter scan over $\lambda$ and $a$, we have constructed a detailed phase diagram (Fig. \ref{fig:PD}) that organizes the distinct shadow morphologies into four well-separated phases.
  Crucially, the formation of cusps, marking the transition between smooth and cuspy shadow boundaries, only becomes possible when $\lambda$ is allowed to vary.
  We have identified a critical value $\lambda_c = \frac{1}{4}(\sqrt{5}-1)$ at which the morphology of the shadow undergoes a qualitative transition. Remarkably, this critical value is independent of both the spin parameter $a$ and the throat radius $r_0$. For $\lambda < \lambda_c$, the outer circular orbits curve and the throat orbits curve meet smoothly with identical slope. No cusp or swallowtail appears.
  For $\lambda > \lambda_c$, the shadow exhibits a prominent cusp with slope discontinuity at the intersection point, clearly visible in Fig. \ref{fig:cross}. The outer unstable orbits curve develops a self-intersecting swallowtail structure.
   Working with fixed $\lambda > \lambda_c$, as the spin parameter is decreased, the two ``ears'' of the swallowtail merge, forming a closed loop; both orbit families still contribute to the boundary.  
   For very small $a$, the throat orbits detach completely from the shadow boundary, rendering the shadow solely determined by the outer unstable circular orbits. 
   Within this phase, for sufficiently small spin (e.g., $a \leq 0.033$ at $\lambda = 2$), the throat orbits further shrink to a single point and ultimately vanish, implying that no light signal from the throat can reach the distant observer.  
The ears-touching boundary exhibits a linear dependence between $a$ and $\lambda$, as shown in Fig. \ref{fig:PD}. The throat-drowning happens only when $a < 0.08$. For higher-spinning wormholes, the throat invariably participates in the shadow. 
  
It is worth emphasizing the separation between the mathematical analysis of the evolution of photon orbits and the actual observable features of the shadow.
Specifically, the observable shadows are the shaded regions in the figures, for example, the shaded region in Fig. \ref{61}.
The cusps themselves are genuine observable features that correspond to points where the tangent direction changes discontinuously, and such sharp corners would be visible in sufficiently high-resolution images. 
In contrast, the swallowtail structure is usually not observable as it is not in the shaded region, and it is a mathematical construct that arises from the parameterization of the photon orbits. 
Although not directly imaged, this structure plays a crucial role in the theoretical analysis.
It allows us to locate the position of the cusp and to calculate the slope of the shadow boundary as one approaches the cusp from either side. 
Similarly, the ears touching behavior describes a stage in the evolution of the unstable circular orbits and does not directly appear in the image. 
However, the throat drowning behavior should be observable because in this regime, the wormhole throat orbits would not contribute to the shadow boundary.
As a result, the overall shape of the shadow changes in a discernible way.
For example, one should be able to tell the difference between the two shadows shown in Figs. \ref{21} and \ref{42}, although they are both smooth.
In the phase diagram shown in Fig. \ref{fig:PD}, the distinctions among the no cusp shadow, cuspy shadow, and throat drowning shadow are expected to be detectable in future high-resolution image observation.

In conclusion, we have demonstrated that the shadow of rotating traversable wormholes exhibits a surprisingly rich structure governed by the redshift parameter $\lambda$ and spin parameter $a$. The emergence of a cusp, the phase diagram comprising four distinct morphologies, and the re-entrant behavior of the throat orbits are novel features characterizing the wormhole shadow.
 This systematic morphological analysis deepens our theoretical understanding of wormhole spacetimes and offers a clean observational diagnostic for constraining $\lambda$ and $a$ with future high-resolution images.

The recent discovery \cite{Wei:2026zwu} shows that cusp formation in black hole shadows constitutes a topological phase transition, characterized by a flip of the topological charge. Moreover, the cusp exhibits the equal-area law and universal critical exponents.
The swallowtail shape of the outer orbits curve and the appearance/disappearance of the cusp for wormhole shadow strongly suggest a similar pattern. 
It is compelling to further study whether the same mechanism from \cite{Wei:2026zwu} applies to the wormhole shadow or not.


\section*{Acknowledgements}
We thank Shao-Wen Wei and Si-Jiang Yang for the helpful discussions. 
We thank the anonymous referee for valuable comments and suggestions.
This work is supported by the National Natural Science Foundation of China (Grant No. 12405073) and the Natural Science Foundation of Tianjin (Grant No. 25JCQNJC01920). PC is partially supported by Tianjin University Self-Innovation Fund Extreme Basic Research Project (Grant No. 2025XJ21-0007).


\appendix
\section{Parameterization of the wormhole throat orbits}\label{sec:para}

For the throat shadow boundary of a rotating wormhole, we can adopt a parameterization of the curve by introducing an effective radius $\bar{r}\in [0,1]$ mimicking the parameterization of outer unstable circular orbits \eqref{eq:ab}.
We can ask the parameterization to satisfy the endpoint conditions and the specified derivative condition. 
The wormhole throat orbits can be written as an ellipse equation
\begin{equation}
	\frac{(\alpha - \alpha_c)^2}{\mathcal{A}^2} + \frac{\beta^2}{\mathcal{A}^2} = 1,
\end{equation}
with ellipse parameters
\begin{equation}
\begin{split}
	\alpha_c &= \frac{\omega_0 r_0^2 \sin \tob}{N_0^2 - \omega_0^2 r_0^2 \sin^2 \tob}, \\
  \mathcal{A} &= \frac{r_0N_0}{N_0^2 - \omega_0^2 r_0^2\sin^2 \tob}, \\
  \mathcal{B} &= \frac{r_0}{\sqrt{N_0^2 - \omega_0^2 r_0^2\sin^2 \tob}}.
\end{split}
\end{equation}
The parametric equations for the throat orbits can be expressed as
\begin{equation}
\begin{aligned}
\alpha(\bar{r}) &= \alpha_c + \mathcal{A} \cdot g(\bar{r}), \\
\beta(\bar{r}) &= \pm\mathcal{B} \sqrt{1 -  g(\bar{r})^2},
\end{aligned}
\end{equation}
with $\bar{r}\in [0,1]$.
$g(\bar{r})$ is the core of the parameterization.
This parameterization should ensure the following properties, which put constraints on $g(\bar{r})$. When $\bar{r} = 0$, it corresponds to the right endpoint of the ellipse $(\alpha_c + \mathcal{A}, 0)$. 
This condition demands $g(0)=1$.
When $\bar{r} = 1$, it corresponds to the impact parameters of the outer unstable orbits at $r_{\text{ph}}\to r_0$, denoted as 
\begin{equation}
\alpha_{\text{ext}} = -\frac{2\lambda}{\omega_0 (3+2\lambda) \sin \theta_{\text{ob}}},\qquad
\beta_{\text{ext}} = \pm \frac{\sqrt{ 9 r_0^2 {\omega_0^2 N_0^{-2}\sin^2\theta_{\text{ob}}} - {4\lambda^2 } }}{(3+2\lambda) \omega_0 \sin\theta_{\text{ob}}} .
\end{equation}
The above condition requires
\begin{equation}
	g(1)=\frac{\alpha_{\text{ext}}-\alpha_c}{\mathcal{A}}\,.
\end{equation}
Moreover, we usually ask the derivatives of $\alpha$ and $\beta$ with respect to $\hat{r}$, at $\hat{r} = 1$, equals derivatives obtained from the outer unstable orbits at $r_{\text{ph}}\to r_0$. This condition further puts constraints on $g'(\bar{r})|_{\bar{r}=1}$.
Once the boundary condition for $\alpha(\bar{r})$ at $\bar{r}=1$ is satisfied, the boundary condition for $\beta(\bar{r})$ would be satisfied automatically, due to the compatibility condition.
One can find a favorite parameterization for the wormhole throat in terms of $g(\bar{r})$, as far as the parameterization is consistent with the boundary conditions for $(\alpha(\bar{r}),\beta(\bar{r}))$.



\begingroup\raggedright\endgroup


\end{document}